\documentclass[twocolumn,prb,floatfix]{revtex4}
\usepackage{graphicx,amsmath}

\begin{document}

\title{Electrostatics of two-dimensional structures: exact solutions and
approximate methods}

\author{Michael M. Fogler}

\affiliation{Department of Physics, University of
California San Diego, La Jolla, California 92093}

\date{\today}

\begin{abstract}

We consider a set of electrostatic problems relevant for determining the
real-space structure and the ground-state energy of a two-dimensional
electron liquid subject to smooth external potentials. Three fundamental
geometries are investigated: an elongated metallic island, an antidot,
and a constriction. In the first two cases complete closed-form
analytical solutions are obtained, despite the absence of rotational or
translational symmetries. These solutions govern the shape and size of
large quantum dots, and also the size of the depletion regions and the
density profiles around isolated antidots. For the constriction, an
exact asymptotical formula for boundary shape is derived and arguments
are given in favor of its universality. For the cases where the full
analytical solution cannot be obtained, an approximate method is
proposed as an alternative. Its accuracy is verified against numerical
simulations in a periodic (checkerboard) geometry.

\end{abstract}
\maketitle

%%%%%%%%%%%%%%%%%%%%%%%%%%%%%%%%%%%%%%%%%%%%%%%%%%%%%%%%%%%%%%%%%%%%%%%%%
\section{Introduction}
\label{Sec:Introduction}

\subsection{Formulation of the problem}

Studies of physical phenomena in two-dimensional metals and thin films
often lead to mixed electrostatic boundary-value
problems.~\cite{Sneddon_book} A prototypical example is the task of
determining the density profile $n(\textbf{r})$ of a two-dimensional (2D)
electron liquid in the proximity of external charges or voltage sources.
Such charges and sources are used in practical applications to
manipulate the electron liquid into desired geometrical shapes, e.g.,
quantum dots, narrow wires, constrictions, \textit{etc\/}. They are also
used to intentionally introduce defects, such as
antidots,~\cite{Deruelle_92,Burnett_93,Goldman_95,Nebel_97} into an
otherwise homogeneous system. In addition to artificial sources of
external potentials, in real materials electrons also experience a
random potential of ubiquitous charged impurities. It appears therefore
that methods that can tackle the corresponding electrostatic problems
could be of considerable value both for applied and for fundamental
research. Unfortunately, the boundary-value problems are notoriously
difficult to solve analytically. The goal of the this paper is to show
that either the complete analytical solutions or their exact asymptotics
can be found for a number of nontrivial basic geometries.

Our most unexpected unexpected result concerns the electrostatics of a
constriction. We show in Sec.~\ref{Sec:Saddle} that the long-range
Coulomb interaction can cause a significant narrowing of the region
occupied by the 2D electron liquid in the constriction and that the
boundaries of this region are described by an interesting
nonanalytic function, see Eq.~(\ref{b}).

%We mention in passing that mathematically similar problems also arise in
%other fields of physics and technology, for instance, in study of
%the electromagnetic field penetration through perforated metallic
%films~\cite{Bethe_44} or scattering of sound and surface acoustic waves
%by apertures in soft screens. However, below we will not pursue those
%problems in any detail and will focus instead on the electrostatics of a
%2D electron liquid.

The present article extends and generalizes the results available in the
literature, e.g., Refs.~\onlinecite{Sneddon_book,Deruelle_92,Burnett_93}
and~\onlinecite{Larkin_92,Chklovskii_92,Chklovskii_93,Fogler_94,%
Davies_94}, and came as an outgrowth of our recent
work~\cite{Fogler_xxx} on the electrostatics of disordered 2D systems.

We will consider 2D electron systems that are separated from grounded
electrodes or other screening bodies by distances much larger than the
interelectron spacing. Such conditions are realized in semiconductor
heterostructures and in field-effect transistors with thick insulator
layers. In these systems electrons interact via the $1/r$ Coulomb law
and form a metallic liquid if their density is not too low.

The results of the present work are obtained within the approximation
that the Thomas-Fermi screening radius $r_\text{TF}$ of the electron liquid
metallic is vanishingly small. This is the correct leading-order
approximation if all lengthscales of interest exceed $r_\text{TF}$.

%Physically, it amounts to neglecting kinetic, exchange, and correlation
%energies in comparison to the energy of the direct Coulomb interaction
%(Hartree term) of the inhomogeneous electron liquid.

In typical semiconductor realizations of 2D systems, $r_\text{TF}$ is of the
order of the interelectron separation $a_\text{e-e} = n^{-1 / 2}$, and so our
approach is good for studying variations of $n(\textbf{r})$ on
lengthscales larger than $a_\text{e-e}$. Note that this does not necessarily
prohibit us from describing some effects that are due to discreteness of
electrons. For example, in Sec.~\ref{Sec:Island} we will be able to
calculate the energy spacing of the Coulomb blockade peaks because in
the leading order it is determined by the classical capacitance. On the
other hand, the equations below cannot be used, e.g., to study quantum
dots with just a few electrons. Also, these equations do not apply too
close to the edges of the metallic regions where $n(\textbf{r}) \to 0$
and so, formally, $a_\text{e-e} \to \infty$. Still, even in these situations
the solution of the corresponding electrostatic problem should provide a
valuable insight.
 
In the approximation of vanishingly small $r_\text{TF}$, the electrostatic
potential $\Phi(\textbf{r})$ in the regions occupied by the electron liquid
is perfectly flat,
\begin{equation}
\text{Metal:}\quad n(\textbf{r}) > 0,
\quad e \Phi(\textbf{r}) = \mu_e = \,\text{const},
\label{bc_metal}
\end{equation}
where $\mu_e$ is the electrochemical potential. The rest of the 2D plane
is occupied by the depletion regions (DR) --- the areas of
exponentially small, effectively zero electron density
that are classically forbidden for the
electrons:
\begin{equation}
\text{DR:}\quad n(\textbf{r}) = 0,
\quad e \Phi(\textbf{r}) > \mu_e.
\label{bc_DR}
\end{equation}
Our goal is to study the conditions that cause DRs to appear and their
detailed stricture.~\cite{Comment_on_nonnegative} To finalize the
formulation of the electrostatic problem we wish to solve for this
purpose, we need an expression for $\Phi$ in terms of $n$. We
distinguish two cases.

\noindent\textit{Case A}.--- If the number of electrons $N_e$ in the system
is finite, we use
\begin{equation}
\Phi(\textbf{r}) = \Phi_{ext}(\textbf{r}) + \frac{e}{\kappa} \int
d^2 r^\prime \frac{n(\textbf{r}^\prime)}{|\textbf{r} - \textbf{r}^\prime|},
\label{Phi_finite_sys}
\end{equation}
where $\Phi_{ext}(\textbf{r})$ is the external electrostatic potential
and $\kappa$ is the dielectric constant of the medium.

\noindent\textit{Case B}.--- If we deal with an infinite system with a
nonzero average electron concentration, we define $\Phi$ as follows:
\begin{equation}
\Phi(\textbf{r}) = \frac{e}{\kappa} \int
d^2 r^\prime \frac{n(\textbf{r}^\prime) - \sigma(\textbf{r}^\prime)}
                  {|\textbf{r} - \textbf{r}^\prime|}.
\label{Phi_inf_sys}
\end{equation}
Here $\sigma(\textbf{r})$ represents a spatially nonuniform background of
opposite charge. In this formulation the primary parameter is the
average background density $n_e = \langle \sigma(\textbf{r}) \rangle$, and
the electrochemical potential in Eqs.~(\ref{bc_metal}) and (\ref{bc_DR})
is determined by electroneutrality of the system as a whole, so that
$\mu_e$ is a function of $n_e$. There is no loss in generality in
assuming that the background charge density $\sigma(\textbf{r})$ is
confined to the same 2D plane as the electrons. Indeed, it is easy to see that
an arbitrary three-dimensional charge density distribution
$\sigma_3(\textbf{r}, z)$ creates essentially the same
electrostatic potential in the $z = 0$ plane as the
following effective 2D density:
\begin{equation}
\sigma(\textbf{r}) = \int d^2 r^\prime \int
\frac{d z |z| \sigma_3(\textbf{r}^\prime, z)}
     {2 \pi [(\textbf{r} - \textbf{r}^\prime)^2 + z^2]^{3/2}}.
\label{sigma_from_sigma_3}
\end{equation}
The only difference between the two potentials is the $\textbf{r}$-independent
term $e^2 C_0^{-1} n_e$, where $C_0^{-1}$, given by
\begin{equation}
C_0^{-1} = \frac{4 \pi}{\kappa}
\frac{\langle \sigma_3 |z| \rangle}{\langle \sigma_3 \rangle},
\label{C_0}
\end{equation}
is the inverse geometric capacitance per unit area between the 2D layer
and the external sources [$\langle \ldots \rangle$ denotes the
three-dimensional (3D) spatial average]. As the name implies, $C_0^{-1}$
is usually determined by the fixed dimensions of the structure and the
electrodes, and so it is almost independent of $n_e$. In this situation,
a finite $C_0^{-1}$ causes only a trivial linear shift of the
electrochemical potential $\mu_e$, while the interesting quantity is the
deviation
\begin{equation}
          \delta\mu \equiv \mu_e - e^2 C_0^{-1} n_e.
\label{delta_mu}
\end{equation}
This is one of the quantities we will be calculating below. We will
assume that $\Phi_{ext}(\textbf{r})$ and $\sigma(\textbf{r})$, whichever is
appropriate, are smooth and bounded functions.

The major difficulty in solving the above equations stems from the mixed
boundary conditions~(\ref{bc_metal}) and (\ref{bc_DR}). These conditions
do not specify where the boundaries of the DRs reside or what the
potential inside the DRs is. They merely state that the DRs may exist
and that inside of them $e\Phi(\textbf{r})$ must exceed a certain constant
value $\mu_e$. It seems that in the general case, one can make only the
following two trivial statements. First, in the
formulation~(\ref{Phi_finite_sys}) the metallic regions are located in
areas where the potential energy $e \Phi_{ext}(\textbf{r})$ is
sufficiently low, the rest of the 2D plane being a DR. Second, in the
formulation~(\ref{Phi_inf_sys}), DRs surround negative local minima of
$\sigma(\textbf{r})$. [If $\sigma$ is nonegative everywhere, then the
sought ground state is simply $n(\textbf{r}) = \sigma(\textbf{r})$ and the
system is free of DRs].

As alluded to above, there has been a sizeable amount of work devoted to
the electrostatics of disordered electron systems. Prominent early
investigations include those of Efros and
Shklovskii~\cite{Efros_Shklovskii_book} on the nonlinear screening in
3D doped semiconductors and its extensions to
2D.~\cite{Gergel_78,Shklovskii_86,Efros_93} In disordered systems DRs
typically have some irregular shapes, and so there is no any convenient
coordinate system that can be used to take advantage of the known
techniques of solving integral equations. Thus, the random case seems
mathematically intractable. Rigorous results that have been obtained for
such problems are limited to certain scaling
laws~\cite{Gergel_78,Shklovskii_86,Efros_93} and some asymptotical
limits.~\cite{Efros_93,Fogler_xxx} There is a hope however that in
regular geometries analytical solutions could be easier to derive. This
is indeed the case. For example, if the external potential
$\Phi_{ext}(x, y)$ depends only on one coordinate, say, $x$, then the
problem can be reduced to an equation for just a few parameters, the
positions of the DR edges.~\cite{Landau_Lifshitz_Elasticity} Once they
are determined, the density $n(x)$ at all other $x$ can be found from
simple analytical formulas, see an example below. Similarly, there is a
closed-form solution in quadratures for the axially-symmetric case,
$\Phi_{ext} = \Phi_{ext}(r)$, provided there is only one
DR.~\cite{Sneddon_book} In other words, it is known how to find the
density profile $n(r)$ of a single round droplet or around a circular
depletion hole. Let us give a few examples.

\subsection{Examples of exact solutions}

\noindent\textit{Example 1}.--- Our first example is a metallic droplet
confined laterally by a parabolic external potential
\begin{equation}
 e \Phi_{ext}(r) = \frac12 U_{x x} r^2,\quad U_{x x} > 0.
\label{Phi_e_parabolic}
\end{equation}
The density profile of such a droplet is known to be
hemispherical~\cite{Sneddon_book}
\begin{equation}
 n(r) = \frac{2}{\pi^2} \frac{\kappa}{e^2} U_{x x} \sqrt{a^2 - r^2},
\label{n_round_droplet}
\end{equation}
with the radius $a$ of the droplet related to its electrochemical
potential by $\mu_e  = U_{x x} a^2$.
%%
%\begin{equation}
% \mu_e = U_{x x} a^2 = \frac{e^2}{\kappa} \left(\frac{9 \pi^2}{16}
%\frac{\kappa}{e^2} U_{x x} N_e^2\right)^{2/3}
%\label{mu_round_droplet}
%\end{equation}
%%
%and $N_e$ is the total number of electrons.~\cite{Fogler_94}

\noindent\textit{Example 2}.--- Another instructive example is an
isolated DR in the form of a perfect circle. As noted above, such a
depletion hole can form when $\sigma(\textbf{r})$ has an isolated negative
minimum, $\sigma_0 < 0$. Let this minimum be located at the origin, $r =
0$, and let the radius $a$ of the induced depletion hole be small enough
so that the expansion
\begin{equation}
 \sigma(\textbf{r}) = \sigma_0 + \frac12 \sigma_{x x} r^2
\label{sigma_round_hole}
\end{equation}
can be used. Following the formalism of Ref.~\onlinecite{Sneddon_book} it
is easy to find that
\begin{equation}
                        a^2 = -3 \sigma_0 / {\sigma_{x x}},
\label{hole_radius}
\end{equation}
so that $n(r) = 0$ at $r \leq a$. At $r > a$, $n(r)$ takes the
form~\cite{Fogler_xxx}
\begin{equation}
n(r) = \frac{\sigma_{x x} a^2}{\pi} \left[
  \sqrt{\frac{r^2}{a^2} - 1}
+ \left(\frac{r^2}{a^2} - \frac{2}{3}\right)
  \arccos \frac{a}{r}\right].
\label{n_round_hole}
\end{equation}

\noindent\textit{Example 3}.--- Our last example is the case of a
$y$-independent background density
\begin{equation}
 \sigma(\textbf{r}) = \sigma(x) = \sigma_0 + \frac12 \sigma_{x x} x^2,
\label{sigma_slit}
\end{equation}
which gives rise to the DR in a form of a stripe of width $2 a$ flanked
by the density distribution~\cite{Landau_Lifshitz_Elasticity}
\begin{eqnarray}
 n(\textbf{r}) &=& \frac12 \sigma_{x x} |x| \sqrt{x^2 - a^2},
\label{n_slit}\\
a^2 &=& -4 \sigma_0 / {\sigma_{x x}}
\label{slit_halfwidth}
\end{eqnarray}
on the two sides. Note that all the presented exact solutions,
Eqs.~(\ref{n_round_droplet}), (\ref{n_round_hole}), and (\ref{n_slit}),
agree with a well-known result~\cite{Landau_Lifshitz_Electrodynamics}
that $n(\textbf{r})$ has a square-root singularity near the edge of the DR.
Previously, these solutions and their generalizations have been used to
study the edges of 2D electron liquid,~\cite{Chklovskii_92} quantum
wires,~\cite{Chklovskii_93,Fogler_94} quantum dots,~\cite{Fogler_94} and
antidots.~\cite{Deruelle_92,Burnett_93}

In a sense, all the aforementioned examples are one-dimensional because
$\Phi_{ext}(\textbf{r})$ [or $\sigma(\textbf{r})$] depend on a single
variable. To the best of our knowledge, there are no published exact
solutions for truly 2D cases, i.e., for the geometries where
$\Phi_{ext}(\textbf{r})$ and $\sigma(\textbf{r})$ are smooth functions
of position and have no translational or rotational symmetries. The
present paper is aimed to fill this gap. The geometries we consider are
as follows. In Sec.~\ref{Sec:Island} we study a droplet in a parabolic
but not necessarily axially-symmetric confining potential. We show that
the droplet has the elliptic shape, with Eq.~(\ref{n_round_droplet})
recovered as a special case. In Sec.~\ref{Sec:Hole} we derive a formula
for the density profile around an elliptic depletion hole. This formula
bridges the limiting cases of Eqs.~(\ref{n_round_hole}) and
(\ref{n_slit}). In Sec.~\ref{Sec:Saddle} we treat the nonlinear
screening problem for the saddle-point, which is presumably the most
interesting basic geometry. We derive the asymptotical formula for the
width of the constriction, Eq.~(\ref{b}), and give arguments in favor of
its universality. In Sec.~\ref{Sec:Chess} we examine the periodic
(``checkerboard'') external potential, which attains two goals. First, it
enables us to study the interplay of the three fundamental geometries
(dot, antidot, and the saddle-point) examined in
Secs.~\ref{Sec:Island}--\ref{Sec:Saddle}. Second, it serves as a
testground for an approximate method of solving electrostatic problems
suggested in our previous publication.~\cite{Fogler_xxx}

At the end of each of the following Sections we briefly comment on the
relevance of the obtained results for various types of experiments. A
detailed comparison with the available experimental data is deferred for
future work.

%%%%%%%%%%%%%%%%%%%%%%%%%%%%%%%%%%%%%%%%%%%%%%%%%%%%%%%%%%%%%%%%%%%%%%%%%
\section{Elliptic island}
\label{Sec:Island}

In this Section we derive the density profile of a metallic
droplet that resides in the external potential
\begin{equation}
 e \Phi_{ext}(\textbf{r}) = \frac12 U_{x x} x^2 + \frac12 U_{y y} y^2,\quad
 0 < U_{x x} \leq U_{y y}.
\label{Phi_e_elliptic_droplet}
\end{equation}
If $U_{x x} = U_{y y}$ we must recover Eq.~(\ref{n_round_droplet}); otherwise,
if $U_{x x} < U_{y y}$, we expect the droplet to be stretched out in the
$x$-direction, along which the confinement is softer.

The quickest way to obtain the solution for arbitrary $U_{x x}$ and
$U_{y y}$ is to use a classic theorem of Frank W.~Dyson.~\cite{Dyson}
One of the corollaries of this theorem concerns the 2D charge density
distribution
\begin{equation}
 n(x, y) = n_d \sqrt{1 - \frac{x^2}{a^2} - \frac{y^2}{b^2}},
\label{n_elliptic_droplet}
\end{equation}
which defines an elliptically shaped droplet with semiaxes $a$ and $b$.
The Dyson theorem indicates that such a droplet creates the in-plane
electrostatic potential of the form
\begin{eqnarray}
 \Phi_d(x, y) &=& ({e}/{\kappa}) \pi n_d
\int\limits_\lambda^\infty \frac{a b\, d l}
                                {\sqrt{(a^2 + l^2) (b^2 + l^2)}}
\nonumber\\
&\times&
\left(1 - \frac{x^2}{a^2 + l^2} - \frac{y^2}{b^2 + l^2}\right),
\label{Phi_elliptic_droplet_I}
\end{eqnarray}
where $\lambda$ is equal to zero inside the droplet and is equal to the
largest root of the equation
\begin{equation}
 \frac{x^2}{a^2 + \lambda^2} + \frac{y^2}{b^2 + \lambda^2} = 1
\label{lambda_elliptic_droplet}
\end{equation}
otherwise. This statement can be proved by expanding $\Phi_d(\textbf{r})$
in ellipsoidal harmonics,~\cite{Comment_on_Dyson} see
Sec.~\ref{Sec:Hole} and App.~\ref{Sec:Ellipsoidal}.

Performing the integration in Eq.~(\ref{Phi_elliptic_droplet_I}) for the case
$\lambda = 0$, we get
\begin{equation}
 \Phi_d = \frac{\pi e b n_d}{\kappa} \Bigl[K
- \frac{K - E}{k^2_d} \frac{x^2}{a^2}
%\nonumber\\
 - \frac{E - (1 - k_d^2) K}{k_d^2} \frac{y^2}{b^2}\Bigr].
\label{Phi_elliptic_droplet_II}
\end{equation}
Here $K$ and $E$ are the complete elliptic integrals of the first and
the second kind, respectively,~\cite{Gradshteyn_Ryzhik} evaluated at
\begin{equation}
                    k_d = \sqrt{1 - ({b}/{a})^2}.
\label{k_d}
\end{equation}
(As explained above, we expect $a \geq b$).
Equations~(\ref{Phi_e_elliptic_droplet}) and
(\ref{Phi_elliptic_droplet_II}) indicate that if we choose $a$, $b$, and
$n_d$ appropriately, we can satisfy the equilibrium
condition~(\ref{bc_metal}). Indeed, we have $e \Phi = e \Phi_{ext} + e
\Phi_d = \mu_e$ in the interior of the droplet if the following
equations hold:
\begin{eqnarray}
\frac{U_{x x}}{U_{y y}} &=&  \frac{(1 - k_d^2)[K(k_d) - E(k_d)]}
                          {E(k_d) - (1 - k_d^2) K(k_d)},
\label{k_d_equation}\\
\frac{a^2}{\mu_e} &=& \frac{2}{U_{x x}} \frac{K(k_d) - E(k_d)}{K(k_d) k_d^2},
\label{a_elliptic_droplet}\\
\frac{n_d^2}{\mu_e} &=& \frac{\kappa^2}{2 \pi^2 e^4}
          \frac{U_{x x} k_d^2}{(1 - k_d^2) (K - E) K}.
\label{n_d_elliptic_droplet}
\end{eqnarray}
It is easy to see that these equations have a unique solution for $a$,
$b$, and $n_d$. Also, from the fact that the integrand in
Eq.~(\ref{Phi_elliptic_droplet_I}) is nonnegative, we conclude
$\Phi_d(\textbf{r})$ decreases with $r$, which ensures that the
inequality~(\ref{bc_DR}) is also satisfied. Thus,
Eq.~(\ref{n_elliptic_droplet}) is the desired solution.

Although the parameters of this solution cannot be expressed in terms of
elementary functions, one can work out the limiting cases, which are as
follows. Let us define
\begin{equation}
                       z_d \equiv U_{y y} / U_{x x} \geq 1.
\label{z_d}
\end{equation}
If $z_d = 1 - \delta z$, where $\delta z \ll 1$, we have a nearly circular
droplet with the following parameters:
\begin{eqnarray}
a^2 &=& \frac{\mu_e}{\sqrt{U_{x x} U_{y y}}}\left[
1 + \frac23 \delta z + O(\delta z^2)\right],
\label{a_nearly_circular_droplet}\\
b^2 &=& \frac{\mu_e}{\sqrt{U_{x x} U_{y y}}}\left[
1 - \frac23 \delta z + O(\delta z^2)\right],
\label{b_nearly_circular_droplet}\\
n_d &=& \frac{2}{\pi^2} \frac{\kappa}{e^2} (U_{x x} U_{y y})^{1/4} \mu_e^{1/2}
 + O(\delta z^2).
\label{n_d_nearly_circular_droplet}
\end{eqnarray}
In the limit $\delta z \to 0$, we indeed recover
Eq.~(\ref{n_round_droplet}). In the opposite limit, $z_d \gg 1$, we have
a strongly elongated droplet with parameters
\begin{eqnarray}
a^2 &=& \frac{2 \mu_e}{U_{x x}} \left[1 - \frac{2}{\cal L} +
O\left(\frac{1}{{\cal L}^2}\right)\right],\quad
{\cal L} \equiv \ln z_d, 
\label{a_elongated_droplet}\\
b^2 &=& \frac{4 \mu_e}{U_{y y}} \left[\frac{1}{\cal L} +
O\left(\frac{1}{{\cal L}^2}\right)\right],
\label{b_elongated_droplet}\\
n_d &\simeq& \frac{1}{2 \pi} \frac{\kappa}{e^2} U_{y y} b
\left(1 + \frac{1}{2 z_d}\right).
\label{n_d_elongated_droplet}
\end{eqnarray}
Note that by setting $z_d$ to infinity and $x$ to zero, we obtain the
solution for another geometry of basic interest: an infinite wire in the
parabolic confining potential $e \Phi_{ext}(y) = U_{y y} y^2 / 2$. The
density distribution in a cross-section of such a wire is a
semicircle,~\cite{Larkin_92}
\begin{equation}
       n(y) = \frac{\kappa}{2 \pi e^2} U_{y y} \sqrt{b^2 - y^2},
\label{n_wire}
\end{equation}
see Eqs.~(\ref{n_elliptic_droplet}) and (\ref{n_d_elongated_droplet}).

Finally, let us calculate the capacitance of the droplet, $C_d = e^2 d
N_e / d \mu_e$, where $N_e$ total number of electrons in the droplet.
Integrating $n$ in Eq.~(\ref{n_elliptic_droplet}) over the area, we get
$N_e = (2 \pi / 3) n_d a b$. Using this result,
Eqs.~(\ref{k_d_equation})--(\ref{n_d_elliptic_droplet}), and some simple
algebra, we find
\begin{equation}
                      C_d = \frac{\kappa}{K(k_d)} a.
\label{C_d}
\end{equation}
As expected, the capacitance scales linearly with the linear size $a$ of
the droplet.

One application of the derived results is the formula for the energy
separation $e \Delta V_g$ of the Coulomb
blockade~\cite{Coulomb_blockade} peaks that would be observed if the
droplet is weakly connected to external leads. Here $V_g$ has the
physical meaning of the voltage on a gate that controls the size of the
droplet, with conversion factor appropriate for the particular
experimental geometry included. In the first
approximation,~\cite{Coulomb_blockade} $e \Delta V_g \simeq e^2 / C_d$.
The analytical asymptotics of this expression are as follows:
\begin{eqnarray}
(e \Delta V_g)^3 &\simeq& \frac{\pi^2}{6}
\frac{e^4}{\kappa^2} \frac{\sqrt{U_{x x} U_{y y}}}{N_e},
\:\:\quad U_{y y} \simeq U_{x x},
\label{DeltaV_nearly_circular_droplet}\\
&\simeq& \frac{1}{12} \frac{e^4}{\kappa^2} \frac{U_{x x}}{N_e}
\ln^2 \frac{U_{y y}}{U_{x x}},
\:\: U_{y y} \gg U_{x x}.
\label{DeltaV_elongated_droplet}
\end{eqnarray}
Thus, for the droplet in a parabolic confinement, the separation between
the Coulomb blockade peaks should scale as $N_e^{-1/3}$, with a
coefficient of proportionality that depends on the asymmetry of the
confining potential. This can be tested in experiments where both the
size and the shape of the quantum dots can be controlled to some degree
independently and the extensive statistics of the Coulomb blockade
spacings can be accumulated, see, e.g., Ref.~\onlinecite{Patel_98}. A
cautionary note is that one should study large dots, $N_e \gg 1$, where
neglected here effects of disorder, single-particle level spacing, or
subtle features of the edge structure~\cite{Ahn_99} (see also
Ref.~\onlinecite{Cohen_99}) are minimal.

%%%%%%%%%%%%%%%%%%%%%%%%%%%%%%%%%%%%%%%%%%%%%%%%%%%%%%%%%%%%%%%%%%%%%%%%%
\section{Elliptic hole}
\label{Sec:Hole}

In this Section we consider the depletion hole that forms in a metallic
2D liquid around an isolated negative minimum of the background charge
density $\sigma(\textbf{r})$. This problem is relevant for, e.g.,
determining the density profile of the electron liquid around an
antidot.

We assume that the minimum of $\sigma(\textbf{r})$ is located at $r = 0$
and that the depletion hole is small enough so that the expansion
\begin{equation}
 \sigma(\textbf{r}) = \sigma_0 + \frac12 \sigma_{x x} x^2
 + \frac12 \sigma_{y y} y^2,\quad 0 < \sigma_{x x} \leq \sigma_{y y},
\label{sigma_elliptic_hole}
\end{equation}
can be used ($\sigma_0 < 0$). If $\sigma_{x x} = \sigma_{y y}$, then DR is a
circle and Eq.~(\ref{n_round_hole}) must be recovered; otherwise, if
$\sigma_{x x} < \sigma_{y y}$, we expect the DR to be elongated in the
$x$-direction, along which $\sigma(\textbf{r})$ has the slowest growth.
Indeed, below we show that the DR is the ellipse
\begin{equation}
S:\:\frac{x^2}{a^2} + \frac{y^2}{b^2} \leq 1,\quad a \geq b.
\label{elliptic_hole_boundary}
\end{equation}
Our solution is based on the following series expansion of the
three-dimensional electrostatic potential $\Phi(\textbf{r}, z)$:
\begin{equation}
 \Phi(\textbf{r}, z) = \sum\limits_{m = 0}^\infty \sum\limits_{\{p\}}
\alpha_m^{p} F_m^p(\lambda) E_m^p(\mu) E_m^p(\nu),
\label{Phi_hole_expansion}
\end{equation}
where $\lambda$, $\mu$, and $\nu$ are the ellipsoidal
coordinates,~\cite{Sneddon_book,Byerly_book}
\begin{equation}
0 \leq \nu^2 \leq k_h^2
\equiv 1 - \frac{b^2}{a^2} \leq \mu^2 \leq 1 \leq \lambda^2
\label{lmn_range}
\end{equation}
that are related to the original Cartesian coordinates $x$, $y$, and
$z$ by
\begin{widetext}
\begin{equation}
x^2 = \frac{a^2}{k_h^2} \lambda^2 \mu^2 \nu^2,
\quad
y^2 = \frac{a^2}{k_h^2 (1 - k_h^2)}
         (\lambda^2 - k_h^2) (\mu^2 - k_h^2) (k_h^2 - \nu^2),
\quad
z^2 = \frac{a^2}{1 - k_h^2}
      (\lambda^2 - 1) (1 - \mu^2) (1 - \nu^2).
\label{lmn}
\end{equation}
The numerical coefficients $\alpha_m^p$ in
Eq.~(\ref{Phi_hole_expansion}) are to be determined, $p$ are certain
real numbers (eigenvalues) that depend on $m$ and $k_h$, and
$E_m^p(\xi)$, $F_m^p(\lambda)$ are the ellipsoidal harmonics of the first
and the second kinds,~\cite{Byerly_book} respectively. The definition
and some properties of $E_m^p$ and $F_m^p$ are reviewed in
App.~\ref{Sec:Ellipsoidal}.

Below we show that only the following $E_m^p$ harmonics~\cite{Byerly_book}
are present in the expansion~(\ref{Phi_hole_expansion}):
\begin{eqnarray}
E_1^{p_1}(\xi) &=& \sqrt{1 - \xi^2},
\label{E_1}\\
E_3^{r_\pm}(\xi) &=& \sqrt{1 - \xi^2}\, (\xi^2 - C_\pm),
\quad
C_\pm = \frac15 \left(1 + 2 k_h^2 \mp \sqrt{1 - k_h^2 + 4 k_h^4}
\right).
\label{E_3}
\end{eqnarray}
(the actual values of $p_1$ and $r_\pm$ will not be needed). As for
$F_m^p$, the corresponding formulas are rather cumbersome for an
arbitrary $0 \leq k_h < 1$. Fortunately, in this Section we will need
primarily their asymptotics at $\lambda^2 \to 1$,
cf.~App.~\ref{Sec:Ellipsoidal}:
\begin{eqnarray}
\frac13 F_1^{p_1}(\lambda) &\simeq& \frac{1}{\sqrt{1 - k_h^2}}
 + D_1\sqrt{\lambda^2 - 1},\quad D_1 = -\frac{E(k_h)}{1 - k_h^2},
\label{F_1_asym}\\
\frac17 F_3^{r_\pm}(\lambda) &\simeq&\frac{(1 - C_\pm)^{-1}}{\sqrt{1 - k_h^2}}
+ D_\pm \sqrt{\lambda^2 - 1},\quad
D_\pm =
\frac{(1 - k_h^2 + 2 C_\pm^2 - 2 C_\pm k_h^2) E(k_h) - (1 - C_\pm)(1 - k_h^2) K(k_h)}
{2 (1 - C_\pm) C_\pm (k_h^2 - C_\pm) (1 - k_h^2)},\quad
\label{F_3_asym}
\end{eqnarray}
\end{widetext}
where $K$ and $E$ are again the complete elliptic
integrals.~\cite{Gradshteyn_Ryzhik}

The series~(\ref{Phi_hole_expansion}) is designed to satisfy the Laplace
equation $(\partial^2 / \partial z^2 + \nabla^2) \Phi(\textbf{r}, z) = 0$
term by term (cf.~Refs.~\onlinecite{Sneddon_book} or
\onlinecite{Byerly_book}). Therefore, our task is to demonstrate that
with a suitable choice of $\alpha_m^p$, $\Phi(\textbf{r}, z)$ also
satisfies the boundary conditions generated by Eqs.~(\ref{bc_metal}),
(\ref{bc_DR}), (\ref{sigma_elliptic_hole}), and
(\ref{elliptic_hole_boundary}):
\begin{eqnarray}
\frac{\partial}{\partial z} \Phi(\textbf{r}, z = \pm 0)
&=& \mp 2 \pi (e / \kappa) \sigma(\textbf{r}),
\quad \textbf{r} \in S,
\label{bc_hole_DR_I}\\
\Phi(\textbf{r}, z = 0) &=& 0,\quad \textbf{r} \not\in S.
\label{bc_hole_metal_I}
\end{eqnarray}
In the last line we set $\mu_e$ to zero, for convenience.

In order to express these boundary conditions in terms of $\lambda$,
$\mu$, and $\nu$ we note that the points immediately above and below the
ellipse $S$ [Eq.~(\ref{elliptic_hole_boundary})] correspond to
$\lambda^2 \to 1 + 0$. Similarly, points immediately above (below) the
rest of the 2D plane correspond to $\mu^2 \to 1 - 0$. The
condition~(\ref{bc_hole_DR_I}) will be satisfied if $\Phi = \Phi_1 +
\Phi_2$, where $\Phi_1$ is analytic in $\lambda^2$ at $\lambda^2 = 1$,
while $\Phi_2$ has a square-root singularity,
\begin{equation}
\Phi_2 \mathop{\sim}_{\lambda^2 \to 1}
 -\frac{2 \pi e a \sigma}{\kappa \sqrt{1 - k_h^2}}
 \sqrt{(\lambda^2 - 1)(1 - \mu^2)(1 - \nu^2)}.\quad
\label{bc_hole_DR_II}
\end{equation}
Expressed as a function of $\mu$ and $\nu$, $\sigma$ takes the form
\begin{eqnarray}
\sigma(\mu, \nu) &=& \sigma_0 + \frac12 \sigma_{y y} a^2 (\mu^2 + \nu^2 - k_h^2)
\nonumber\\
&+& \frac12 \frac{a^2}{k_h^2} (\sigma_{x x} - \sigma_{y y}) \mu^2 \nu^2.
\label{sigma_mu_nu}
\end{eqnarray}
Finally, Eq.~(\ref{bc_hole_metal_I}) is equivalent to
\begin{equation}
\Phi(\lambda, \mu = 1, \nu) = 0.
\label{bc_hole_metal_II}
\end{equation}
A quick examination of Eqs.~(\ref{bc_hole_DR_II}) and
(\ref{sigma_mu_nu}) makes the above claim that the
series~(\ref{Phi_hole_expansion}) involves only the ellipsoidal
harmonics defined by Eqs.~(\ref{E_1})--(\ref{F_3_asym}) plausible.
Assuming that this is true, we conclude that on the ellipse $S$, $\Phi$
must factorize as follows:
\begin{equation}
\Phi = \sqrt{(1 - \mu^2)(1 - \nu^2)}\, P(\mu^2, \nu^2),
\label{Phi_hole_DR_I}
\end{equation}
where $P$ is a polynomial of the second degree symmetric in its two
arguments. Since $\Phi$ must satisfy Eq.~(\ref{bc_hole_metal_II}), it
has to be of the form
\begin{equation}
\Phi = \frac{\Phi_0}{(1 - k_h^2)^{3/2}} (1 - \mu^2)^{3/2}
 (1 - \nu^2)^{3/2},\quad \lambda = 1,
\label{Phi_hole_DR_II}
\end{equation}
where $\Phi_0$ is some constant. Combined with the expressions for
$F_m^p(1)$ that follow from Eqs.~(\ref{F_1_asym}) and (\ref{F_3_asym}), this
fixes the expansion coefficients $\alpha_m^p$ to be
\begin{eqnarray}
  \alpha_1^{p_1} &=& -\frac{1}{5} \Phi_0,
\label{alpha_p_1}\\
  \alpha_3^{r_-} &=& -\alpha_3^{r_+} = \frac{3}{14}
  \frac{\Phi_0}{\sqrt{1 - k_h^2 + 4 k_h^4}}.
\label{alpha_r_-}
\end{eqnarray}
After some more algebra one finds that the boundary
condition~(\ref{bc_hole_DR_II}) can indeed be satisfied if
$k_h$ is the solution of the equation
\begin{equation}
\frac{\sigma_{x x}}{\sigma_{y y}} = (1 - k_h^2)
\frac{(2 k_h^2 - 1) E(k_h) + (1 - k_h^2) K(k_h)}
     {(1 + k_h^2) E(k_h) - (1 - k_h^2) K(k_h)},
\label{k_h_equation}
\end{equation}
while $a$ is set by
\begin{equation}
a^2 = -\frac{2 \sigma_0}{\sigma_{y y}}
\frac{(1 + k_h^2) E(k_h) - (1 - k_h^2) K(k_h)}{k_h^2 (1 - k_h^2) E(k_h)}.
\label{a_elliptic_hole}
\end{equation}
For the potential $\Phi_0$ at the center of the DR we get
\begin{equation}
\Phi_0 = \frac{2 \pi}{3} \frac{e}{\kappa}
\frac{k_h^2 (1 - k_h^2)^{3/2}}
     {(1 + k_h^2) E - (1 - k_h^2) K} \sigma_{y y} a^3,
\label{Phi_0_hole}
\end{equation}
while at other points on $S$, $\Phi(\textbf{r})$ takes the form
stipulated by Eq.~(\ref{Phi_hole_DR_II}):
\begin{equation}
\Phi(\textbf{r}) = \Phi_0
\left(1 - \frac{x^2}{a^2} - \frac{y^2}{b^2}\right)^{3/2}
\quad (\textbf{r} \in S).
\label{Phi_hole_DR_III}
\end{equation}
At large distances from the origin, $\Phi(\textbf{r}, z)$ is
dominated by the $m = 1$, $p = p_1$ term in Eq.~(\ref{Phi_hole_expansion})
and behaves similar to the potential of an electric dipole,
\begin{equation}
\Phi(\textbf{r}, z) \sim \frac{p_h}{\kappa} \frac{|z|}{(r^2 + z^2)^{3 / 2}},
\quad p_h = \frac{4 \pi}{15} \frac{e}{E} \sigma_0 a b^2,
\label{Phi_hole_far}
\end{equation}
except the apparent dipole moment $\pm p_h \hat{\bf z}$ has the
opposite direction at observation points above and below the $z = 0$
plane. Our formula for $p_h$ resembles a well known
result~\cite{Landau_Lifshitz_Electrodynamics,Bethe_44} for the apparent
dipole moment of a round hole in a metallic sheet.

The sought density distribution $n(\textbf{r})$ outside of $S$ is given by
\begin{equation}
n(\textbf{r}) = \sigma(\textbf{r}) + \frac{\kappa}{2 \pi e}
             \lim_{z \to 0} \frac{\Phi}{|z|},
\label{n_hole_from_Phi_I}
\end{equation}
or, equivalentely, by
\begin{equation}
n = \sigma + \frac{\kappa}{2 \pi e a} \lim_{\mu^2 \to 1} 
 \frac{\Phi(\lambda, \mu, \nu)  \sqrt{1 - k_h^2}}
      {\sqrt{(\lambda^2 - 1)(1 - \mu^2)(1 - \nu^2)}}.
\label{n_hole_from_Phi_II}
\end{equation}
At large $r$ we can use Eq.~(\ref{Phi_hole_far}) to obtain
\begin{equation}
   n(\textbf{r}) \simeq \sigma(\textbf{r})
+ \frac{p_h}{2 \pi e} \frac{1}{r^3},\quad r \gg a,
\label{n_hole_far}
\end{equation}
which elucidates how the approach to the perfect screening, $n(\textbf{r})
\to \sigma(\textbf{r})$, takes place away from the DR. As for $n(\textbf{r})$
near the DR, it is rather complicated, except in two cases: (1) a round
depletion hole and (2) a slit infinite in the $x$-direction. In the
first case the solution is given by Eq.~(\ref{n_round_hole}) (see
App.~\ref{Sec:Ellipsoidal} for details); in the second case, realized at
$\sigma_{x x} = 0$, the solution is given by Eq.~(\ref{n_slit}) upon
replacements $x \to y$, $a \to b$.

It is also possible to work out analytically the lowest-order
corrections to these two limiting cases. Thus, a highly asymmetric
$\sigma$-minimum, $z_h \equiv \sigma_{x x} / \sigma_{y y} \ll 1$, gives
rise to a strongly elongated DR with parameters
\begin{eqnarray}
  a^2 &=& -\frac{2 \sigma_0}{\sigma_{x x}}
        \left[1 + {\cal L}{z_h} + o(z_h)\right],
\quad {\cal L} \equiv \ln z_h^{-1} - 1,\quad\quad
\label{a_elongated_hole}\\
  b^2 &=& -\frac{4 \sigma_0}{\sigma_{y y}}
        \left[1 - \frac12 {\cal L}{z_h} + o(z_h)\right],
\label{b_elongated_hole}\\
\Phi_0 &=& -\frac{8 \pi}{3} \frac{e}{\kappa}
            \frac{|\sigma_0|^{3 / 2}}{\sigma_{y y}^{1 / 2}}
        \left[1 - \frac12 {\cal L}{z_h} + o(z_h)\right].
\label{Phi_0_elongated_hole}
\end{eqnarray}
Equation~(\ref{slit_halfwidth}) is recovered in the limit $z_h \to 0$.

On the other hand, if the $\sigma$-minimum is nearly axially symmetric,
$\delta z \equiv 1 - z_h \ll 1$, then the corresponding DR is nearly
circular, with the parameters
\begin{eqnarray}
  a^2 &=& -\frac{3 \sigma_0}{\sqrt{\sigma_{x x} \sigma_{y y}}}
        \left[1 + \frac25 \delta z + O(\delta z^2)\right],
\label{a_nearly_circular_hole}\\
  b^2 &=& -\frac{3 \sigma_0}{\sqrt{\sigma_{x x} \sigma_{y y}}}
        \left[1 - \frac25 \delta z + O(\delta z^2)\right],
\label{b_nearly_circular_hole}\\
\Phi_0 &=& -\frac{8}{\sqrt{3}} \frac{e}{\kappa}
            \frac{|\sigma_0|^{3 / 2}}{(\sigma_{x x} \sigma_{y y})^{1 / 4}}
        + O(\delta z^2).
\label{Phi_0_nearly_circular_hole}
\end{eqnarray}
At $\delta z = 0$, Eqs.~(\ref{a_nearly_circular_hole}) and
(\ref{b_nearly_circular_hole}) reduce to Eq.~(\ref{hole_radius}).

Let us now discuss the effect of an isolated elliptic DR on the total
energy of the system. It is easy to see that the DR causes a positive
correction $\Delta E$ to the energy, which can be estimated as follows. The
unscreened electric field that surrounds the DR is of the order of
${\cal E} \sim 2 \pi (e / \kappa) \sigma_0$ and is concentrated mainly
in a volume $\Delta V \sim a \times b \times b$. Hence,
\begin{equation}
\Delta E = \frac{\kappa}{8 \pi} \int d^2 r \int d z
 {\cal E}^2(\textbf{r}, z) \sim \frac{e^2}{\kappa} \sigma_0^2 a b^2.
\label{Delta_E_estimate}
\end{equation}
The exact calculation of $\Delta E$ is more convenient to perform in terms of the
potential $\Phi(\textbf{r})$ and the total charge density $e n(\textbf{r}) - e
\sigma(\textbf{r})$,
\begin{eqnarray}
\Delta E &=& \frac12 \int d^2 r \Phi(\textbf{r}) e [n(\textbf{r})
 - \sigma(\textbf{r})]\nonumber\\
 &=& \frac12 e \Phi_0 \int\limits_S d^2 r \sigma(\textbf{r})
\left(1 - \frac{x^2}{a^2} - \frac{y^2}{b^2}\right)^{3/2}
\nonumber\\
 &=& \frac{16 \pi^2}{105} \frac{e^2}{\kappa}
     \frac{\sigma_0^2 a b^2}{E(k_h)}.
\label{Delta_E}
\end{eqnarray}
The result is in agreement with the preceding estimate. [Recall that $1
\leq E(k_h) \leq \pi / 2$, and so $E(k_h)$ does not have a strong
dependence on $a / b$.\/] For fixed $\sigma_{x x}$ and $\sigma_{y y}$,
we have $a, b \propto |\sigma_0|^{1 / 2}$, so that $\Delta E \propto
|\sigma_0|^{7/2}$.

The energy correction due to DRs affects a number of experimentally
measurable properties of 2D electron systems. One example is the
magnetization of quantum dots and quantum wires under the quantum Hall
effect conditions. Although the DRs with $n(\textbf{r}) = 0$ do not appear
in that context, a very similar role is played by regions of locally
depleted topmost Landau level.~\cite{Burnett_93,Fogler_94}.

\begin{figure}
\centerline{
\includegraphics[width=2.1in]{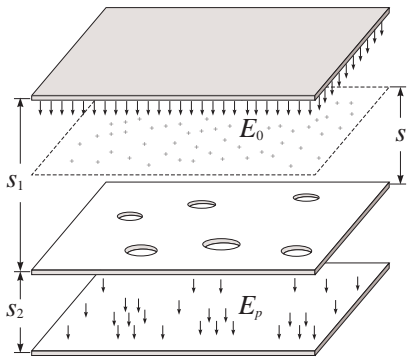}
}
\vspace{0.2in}
\setlength{\columnwidth}{3.2in}
%\centerline{
\caption{
The geometry of the field-penetration experiment. The 2D layer of
interest is sandwiched between the top and the bottom metallic gates.
The sample contains ionized impurities (dopants) that are shown as
randomly scattered plus signs. The case of $\delta$-doping is assumed
where all the dopants reside in a plane parallel to the 2D layer. The
impurities create a random electric field that nucleates the depletion
regions (shown as holes in the probed layer). These depletion regions
enhance the penetrating electric field $E_p$ that can be detected with
the help of the bottom gate.
\label{Fig_field_penetration}
}
%}
\end{figure}

Another example concerns the DRs induced by random impurities, see
Fig.~\ref{Fig_field_penetration}. Such DRs can strongly influence the
energy density of macroscopic 2D systems, especially at low electron
densities. This effect may be important~\cite{Fogler_xxx} for the
actively studied phenomenon of the 2D metal-insulator
transition.~\cite{Abrahams_01}

To show how the obtained formulas can be applied in this context let us
discuss one experimental technique, which is a particular sensitive
probe of DRs. This technique, pioneered by
Eisenstein,~\cite{Eisenstein_94} is the measurement of the electric
field penetration. In order to do this type of
experiment,~\cite{Eisenstein_94,Dultz_00,Ilani_00} one prepares a
structure where the 2D layer is sandwiched between two electrodes that
can be considered good metals. Once a voltage difference is applied
between the 2D layer and the top electrode, some electric field leaks
through the 2D layer and its spatial average $\langle E_p \rangle$ can
be measured by monitoring the amount of electric charge that has flown
into the bottom layer, see Fig.~\ref{Fig_field_penetration}. It is
immediately obvious that DRs must enhance $\langle E_p \rangle$. Using
the results of this Section, we are now able to calculate how this
enhancement of $\langle E_p \rangle$ is related to the concentration and
the linear sizes of the DRs. In this calculation we will assume that the
interlayer distances $s_1$ and $s_2$ (see
Fig.~\ref{Fig_field_penetration}) are large.

%
% FIG. 1
%

Equation~(\ref{Phi_hole_far}) gives one way to calculate $\langle E_p
\rangle$. An alternative and a more general way is to work with the
energy correction $\Delta E$ and the corresponding contribution $\Delta
C^{-1}$ to the inverse capacitance per unit area of the device. [$\Delta
C^{-1}$ is the correction to the inverse geometric capacitance $C_0^{-1}
= 4 \pi s_1 / \kappa$, cf.~Eq.~(\ref{C_0})]. For a single elliptic DR
$\Delta E$ and $\Delta C^{-1}$ are related by
\begin{equation}
\Delta C^{-1} = \frac{A}{e^{2}} \frac{d^2 \Delta E}{d N_e^2}
 = \frac{1}{e^{2} A} \frac{d^2 \Delta E}{d \sigma_0^2} = 
\frac{4 \pi^2}{3 \kappa A} \frac{a b^2}{E(k_h)},
\label{Delta_C_hole}
\end{equation}
where $A$ is the sample area. If the DRs in the 2D layer are well
separated, their contributions to $\Delta C^{-1}$ are, in the first
approximation, additive. The total correction to the inverse
capacitance, $\Delta C^{-1}_{tot}$, characterizes the nonideal nature of
our capacitor and therefore the amount of field penetration. Indeed, a
simple derivation along the lines of Ref.~\onlinecite{Eisenstein_94}
leads to the relation
\begin{equation}
\frac{d}{d n_e} \langle E_p \rangle = \frac{e}{s_2} \Delta C^{-1}_{tot}
= \frac{4 \pi^2}{3} \frac{e}{\kappa} \frac{N_\text{DR}}{s_2}
\left\langle \frac{a b^2}{E(k_h)}\right\rangle,
\label{Delta_C_from_E_p}
\end{equation}
where $N_\text{DR}$ is the DR concentration.

Note that as all the other results in this paper,
Eq.~(\ref{Delta_C_from_E_p}) is derived in the approximation that
ignores non-Hartree terms in the energy (kinetic and
exchange-correlation energies). This is not a serious omission at low
electron densities $n_e$ where the DRs dominate the field penetration.
However, at higher densities, the non-Hartree terms must be included.
It may lead to an interesting nonmonotonic behavior of $\Delta
C^{-1}_{tot}(n_e)$, discussed theoretically in
Refs.~\onlinecite{Efros_93} and~\onlinecite{Fogler_xxx}.

%The results of this Section and, in particular, Eq.~(\ref{Delta_C_hole})
%may also be relevant for understanding intriguing Coulomb blockade-like
%phenomena~\cite{Coulomb_blockade} that occur around an antidot in the
%presence of a quantizing magnetic field normal to the 2D
%plane.~\cite{Goldman_95,Kataoka_99} We hope to address this
%problem in a future publication.

%%%%%%%%%%%%%%%%%%%%%%%%%%%%%%%%%%%%%%%%%%%%%%%%%%%%%%%%%%%%%%%%%%%%%%%%%
\section{Saddle-point}
\label{Sec:Saddle}

In this Section we examine the structure of the 2D electron liquid near
a saddle-point of the external potential,
\begin{equation}
\Phi_{ext} = \frac12 U_{y y} y^2 - \frac12 U_{x x} x^2 + O(r^4),
\quad U_{x x}, U_{y y} > 0.
\label{Phi_saddle}
\end{equation}
This is the most interesting fundamental geometry but also the most
difficult one for the analytical study.

It is easy to understand that the electron liquid should be confined in
the $y$-direction into a strip that has the smallest width at $x = 0$
and widens up as $|x|$ increases. In other words, this structure can be
thought of as a local constriction in a 2D electron system.
Such structures have been fabricated and intensively studied
in the past. Especially much attention has been devoted to quantum point
contacts, which are constrictions with the bottleneck width
comparable to the interelectron separation.~\cite{van_Wees_88}
Nevertheless, to the best of out knowledge, the analytic solution for
the electron density profile $n(x, y)$ and the total electrostatic
potential $\Phi(x, y)$ around the constriction has not been presented.
(For a recent numerical work, see Ref.~\onlinecite{Tkachenko_01}).
Instead, the unknown functional forms of $n$ and $\Phi$ have been
approximated by expressions chosen somewhat arbitrarily. For $\Phi({\bf
r})$, a parabolic function~\cite{Fertig_87,Buttiker_90} [as in
Eq.~(\ref{Phi_saddle})] or a parabola with a flat insert~\cite{Frost_94}
have been used. The boundary of the 2D electron channel was modelled
variously by combination of wedges,~\cite{Szafer_89} circular
arcs,~\cite{He_93} confocal hyperbolas, or more complicated
curves.~\cite{Maao_94}

Let $y = \pm b(x)$ be the equation for the boundaries of the electron
liquid in the constriction. Although we have not succeeded in solving
the electrostatics problem for the saddle-point completely, below we
show that the correct asymptotical behavior of $b(x)$ is given by
\begin{equation}
        b(x) = \alpha_0 |x| \exp
        \left(-\sqrt{\ln \left|\frac{x_0}{x}\right|}\,\,\right),
        \quad |x| \ll x_0,
\label{b}
\end{equation}
where $\alpha_0$, $x_0$ are constants determined by the boundary
conditions far from the saddle-point. This is perhaps the most
interesting theoretical result that we achieve in this work.

The exponential factor in Eq.~(\ref{b}) is a nontrivial effect of the
long-range Coulomb interaction (see below). It is therefore interesting
to do a quick estimate of this factor for a typical experimental setup.
As explained in Sec.~\ref{Sec:Introduction}, pure electrostatics is not
adequate on distances shorter than the interparticle separation
$a_\text{e-e}$. Therefore, in reality, Eq.~(\ref{b}) applies only as long as
$b(x), |x| \gg a_\text{e-e} \sim 1 / \sqrt{n(x, 0)}$. On the other hand,
$x_0$ can be large. For example, using $x_0 \sim 1 \mu{m}$ and $x =
a_\text{e-e} \sim 50\,nm$, which are reasonable numbers for GaAs-based point
contacts, we obtain $\ln(x_0 / x) \sim 3.0$, so that the exponential
factor in Eq.~(\ref{b}) is approximately $0.2$. Thus, the long-range
interaction can cause a significant narrowing of the bottleneck of the
constriction.

In the purely electrostatic model ($a_\text{e-e} = 0$) we are allowed to
consider the limit $x \to 0$. Then Eq.~(\ref{b}) predicts that the
tangent $\alpha(x) = b(x) / x$ of the opening angle of the constriction
becomes vanishingly small, independently of $U_{y y} / U_{x x}$. This
behavior stems from the long-range nature of the Coulomb interaction
between the electrons. With short-range interactions (or without
interactions) electrons fill the constriction up to the equipotential
contour $\Phi_{ext}(x, y) = \mu_e$; therefore, the constriction is
either bounded by two confocal hyperbolas or, when $b(0) \to 0$, by the
two straight lines. In the latter case $\alpha = (U_{x x} / U_{y
y})^{1/2}$, which is finite, in contrast to Eq.~(\ref{b}). The physical
reason for the difference is the ability of the system with long-range
interactions to modify (screen) the external potential even at points
$\textbf{r} = (x, y)$ that are \textit{outside\/} of the area occupied
by the electrons. The screening flattens out the confining potential
$\Phi_{ext}$, allowing the electron liquid to spread over a larger area
compared to the noninteracting limit. The nontrivial statement embodied
by Eq.~(\ref{b}) is that the screening is more effective along the
longitudinal ($x$) direction, so that the extra area occupied by the
interacting electrons is strongly funneled into the constriction center.

Let us proceed to the derivation of Eq.~(\ref{b}). We distinguish two
cases, which are treated in two separate subsections below.

\subsection{Small opening-angle constriction}

First we will treat the limit $U_{x x} \ll U_{y y}$ where $\Phi_{ext}$
is a slow function of $x$. In this case the constriction can be thought
of as an adiabatically narrowing quantum wire.~\cite{Glazman_88} To the
lowest order in parameter $U_{x x} / U_{y y}$, the $y$-dependence of
$n(x, y)$ at a given fixed $x$ can be found by solving the requisite
electrostatic problem assuming that $\Phi_{ext}$ is $x$-independent. The
solution for $n(\textbf{r})$ is given by Eq.~(\ref{n_wire}) where $b$
should now be understood as a yet to be found function of $x$. The
electrostatic potential is also easy to find,
\begin{equation}
\Phi(\textbf{r}) = \mu_e + \frac12 U_{y y}
 \left(y \sqrt{y^2 - b^2} - b^2 \cosh^{-1} \frac{y}{b} \right).
\label{Phi_wire}
\end{equation}
Let us define the one-dimensional (1D) electron density $q(x)$ along the
wire by
\begin{equation}
 q(x) = \int\limits_{-b}^{b} d y\, n(x, y)
      = \frac{1}{4} \frac{\kappa}{e^2} U_{y y} x^2 \alpha^2(x).
\label{q_wire}
\end{equation}
In view of the last equation, the desired $b(x)$ is directly related to
$q(x)$. To find the equation for $q(x)$ we substitute Eq.~(\ref{n_wire})
into the boundary condition~(\ref{bc_metal}) and take $y = 0$. This
yields
\begin{equation}
\int\limits_{-x_0}^{x_0} d x^\prime G(x, x^\prime) q(x^\prime)
 = \frac12 U_{x x} x^2 - \mu_e - \Delta\Phi(x_0).
\label{Eq_for_q}
\end{equation}
The kernel $G(x, x^\prime)$ can be expressed in terms of elliptic
integrals. At $|x - x^\prime| \gg b(x^\prime)$, it reduces to the
Coulomb interaction potential $G(x, x^\prime) \simeq e^2 / \kappa |x -
x^\prime|$. The term $\Delta\Phi(x_0)$ in Eq.~(\ref{Eq_for_q})
represents the potential created by electrons located at points $|x| >
x_0$. It is determined by the behavior of $\Phi_{ext}$ at large
distances and therefore high energies. On the other hand, we are
interested mainly in the structure of the constriction as small $x$.
Physically, we may expect that minor changes in $\mu_e$ modify $b(x)$ near
the origin considerably but leave $\Delta\Phi(x_0)$ virtually the same.
Therefore, the role of $\Delta\Phi(x_0)$ is simply to renormalize the
electrochemical potential by a constant: $\mu_e \to \mu_e +
\Delta\Phi(x_0)$.

The idea of renormalization proves to be very fruitful in the present
problem. Indeed, since we are interested in the behavior at small
distances, we have a freedom in choosing the cutoff as long as it
exceeds the distance of interest. Then $\mu_e$ should be viewed as a
function of the cutoff and as such, it must satisfy a certain
renormalization group (RG) equation. Similarly, there must be an RG
equation for $U_{x x}$ and, in fact, for $U_{y y}$. From
Eq.~(\ref{Eq_for_q}) one finds that to the lowest order in the parameter
$\alpha(l) \ll 1$, these equations are
\begin{equation}
 \frac{d}{d l} U_{x x} = - U_{y y} \alpha^2(l),\quad
 \frac{d}{d l} U_{y y} = - \frac12 U_{y y} \alpha^2(l),
\label{RG_Eqs}
\end{equation}
where $l = \ln |x_0 / x|$, $x$ being the running cutoff. The RG flow
persists as long as~\cite{Comment_on_cutoff} $|x| \gg b(x)$. Let us
consider the most interesting case, $b(0) = 0$, where $|x|$ is always
larger than $b$, so that the RG is able to reach its fixed point.
What is this fixed point? To the order we are working with, $U_{y y} =
U_{y y}(x_0) + [U_{x x} - U_{x x}(x_0)] / 2$. This implies that for
$U_{y y}(x_0) \gg  U_{x x}(x_0)$ the renormalization of $U_{y y}$ can
be ignored. Therefore, at the fixed point, where the right-hand sides of
Eq.~(\ref{RG_Eqs}) must vanish, $\alpha(l = \infty) = \alpha(x = 0) =
0$, in agreement with Eq.~(\ref{b}) and statements above.

Let us now derive the complete form of $\alpha(x)$. From Eqs.~(\ref{q_wire})
and (\ref{Eq_for_q}) we find that
\begin{equation}
 -\alpha^2(l) [\ln \alpha(l) + O(1)] + \int\limits_{0}^{l}
 d l^\prime \alpha^2(l^\prime)
 = \frac{U_{x x}(x_0)}{U_{y y}}.
\label{Eq_for_alpha}
\end{equation}
Differentiating both sides with respect to $l$, we get
\begin{equation}
\frac{d}{d l} \alpha \simeq \frac{\alpha}{2 \ln \alpha},
\label{RG_for_alpha}
\end{equation}
which complements Eq.~(\ref{RG_Eqs}) and has~(\ref{b}) as the solution.

It is instructive to rewrite Eq.~(\ref{b}) in terms of the renormalized
$U_{x x}$ [this can be done by integrating Eq.~(\ref{RG_Eqs})]:
\begin{equation}
\displaystyle \alpha^2(l) = 2 \frac{U_{x x}(l)}{U_{y y}}
\left[1 + \ln \frac{U_{x x}(0)}{U_{x x}(l)}\right]^{-1}.
\label{alpha_RG}
\end{equation}
Since $U_{x x}(l = \infty) = 0$ [see Eqs.~(\ref{RG_Eqs}) and
(\ref{Eq_for_alpha})], Eq.~(\ref{alpha_RG}) describes the decrease of
$\alpha(x)$ from its noninteracting limit value
at $x = x_0$ to the asymptotical zero at $x = 0$.

The obtained RG fixed point must have a finite basin of attraction,
presumably $\alpha \lesssim 1$, for which Eq.~(\ref{b}) must be valid.
The strong evidence that Eq.~(\ref{b}) is the universal asymptotical law
independently of the starting $\alpha(x_0)$ is furnished by the
following analysis of the case $\alpha(x_0) \gg 1$, which is the far
departure from the found fixed point.

\subsection{Large opening-angle constriction}

If the starting, i.e., long-distance value of $\alpha$ is large, it is
no longer convenient to parametrize the saddle-point by $U_{x x}$ and
$U_{y y}$. Indeed, now the external potential $\Phi_{ext}$ is almost
completely screened, and so it is rather useless in setting up the
problem. Instead, as in Sec.~\ref{Sec:Hole}, we describe the effect of
the external sources by an effective neutralizing charge density
$\sigma$, so that the total charge density in the system is
$n(\textbf{r}) - \sigma(\textbf{r})$, see Sec.~\ref{Sec:Introduction}.
Near the origin, $\sigma(\textbf{r})$ should be of a saddle-point type,
\begin{equation}
\sigma = \sigma_0 + \frac12 \sigma_{x x} x^2
 - \frac12 \sigma_{y y} y^2,
\quad 0 < \sigma_{y y} \ll \sigma_{x x}.
\label{sigma}
\end{equation}
In fact, working with $\sigma(\textbf{r})$ instead of $U(\textbf{r})$
also brings us closer to the practical side of fabrication of such a
large-angle constriction. It seems feasible that one can make this
type of a structure by depositing a static surface charge of the
form~(\ref{sigma}) nearby the 2D plane but doing it with voltage sources
(thin metallic gates) may be problematic. Although in the former method
$\sigma_{x x}$ and $\sigma_{y y}$ in Eq.~(\ref{sigma}) would likely be
fixed once the structure is made, $\sigma_0$ could still be varied by an
additional distant gate on top of the device.

As in the case of the small-angle constriction, the relation between
$\sigma_0$ and $\mu_e$, i.e., the top-gate voltage in the suggested
setup, is determined by behavior at large distances. Consequently, in a
small interval of $\mu_e$ of interest to us we should have
\begin{equation}
                   \sigma_0 = ({C} / {e^2}) \mu_e,
\label{sigma_0}
\end{equation}
where $C$ is a constant approximately equal to $C_0$, the geometric
capacitance per unit area [cf.~Eq.~(\ref{C_0})].

We have a situation where the 2D plane is almost completely covered by
the metallic liquid, except a narrow depletion strip that gradually fans
out of the origin along the $y$-axis. This problem is adiabatic with
respect to coordinate $y$. To the lowest order in $1 / \alpha_1$, where
\begin{equation}
          \alpha_1 = (\sigma_{x x} / 2 \sigma_{y y})^{1 / 2}, 
\label{alpha_1}
\end{equation}
we can solve the system of Eqs.~(\ref{bc_metal}), (\ref{bc_DR}),
(\ref{Phi_inf_sys}), and (\ref{sigma}) pretending that $\sigma$ is
$y$-independent. The solution is an obvious modification of
Eqs.~(\ref{n_slit}) and (\ref{Phi_hole_DR_III}):
\begin{eqnarray}
 n(x, y) &=& \frac12 \sigma_{x x} |x| \sqrt{x^2 - a^2(y)},
\label{n_slit_saddle}\\
a^2(y) &=& -4 \sigma(0, y) / \sigma_{x x},
\label{w_perp}\\
\Phi(x, y) &=& \mu_e + \frac{\pi}{3} \frac{e^2}{\kappa} \sigma_{x x}
  [a^2(y) - x^2]^{3 / 2}.
\label{Phi_slit}
\end{eqnarray}
These equations apply whenever they give real $n$ and $\Phi$; otherwise,
$n = 0$ and $\Phi = \mu_e$. At this level of aproximation, the
boundaries of the constriction are confocal hyperbolas defined by the
equation $x^2 = a^2(y)$, see Eq.~(\ref{w_perp}). At $\sigma_0 = 0$ where
the constriction just opens up, these hyperbolas become straight lines
with $\alpha = \alpha_1 = \text{const}$, while $\Phi$ acquires the cubic
dependence on $y$:
\begin{equation}
\Phi(0, y) = \frac{\pi}{24} \frac{e^2}{\kappa}
\sigma_{y y}^{3 / 2} \sigma_{x x}^{-1 / 2} |y|^3.
\label{Phi_y_wide}
\end{equation}
Let us now show that in a more careful treatment,
Eqs.~(\ref{n_slit_saddle})--(\ref{Phi_y_wide}) become invalid at
exponentially small $x$ and $y$. It is convenient to assume that at
large $y$, $\sigma(\textbf{r})$ changes from the decreasing to an
increasing function of $y$ so that the depletion region terminates at
some $|y| = y_0$. In this case the system can be thought of as a 2D
metallic sheet with an elongated bowtie-shaped hole. Similar to the case
of a round hole,~\cite{Jackson_book} the 3D electrostatic potential at
$|x|, |z| \gg a(y)$ can be sought in the form
\begin{equation}
\Phi(\textbf{r}, z) - \mu_e \simeq \frac{e^2}{\kappa}
\int\limits_{-y_0}^{y_0}
\frac{d y^\prime p(y^\prime) |z|}
     {[x^2 + z^2 + (y - y^\prime)^2]^{3 / 2}},
\label{Phi_far}
\end{equation}
which is the lowest-order (dipolar) term in the multipole expansion
compatible with Eq.~(\ref{bc_metal}). Since the total
charge density $\Delta n(\textbf{r}) = n(\textbf{r}) - \sigma(\textbf{r})$ is
proportional to the discontinuity in $\partial \Phi / \partial z$ at $z
= 0$, Eq.~(\ref{Phi_far}) entails
\begin{equation}
\Delta n(\textbf{r}) \simeq \frac{1}{2 \pi}
\int\limits_{-y_0}^{y_0}
\frac{d y^\prime p(y^\prime)}
     {[x^2 + (y - y^\prime)^2]^{3 / 2}}.
\label{n_far}
\end{equation}
To find $p(y)$ we match the near and far-field asymptotics,
Eqs.~(\ref{n_slit}) and (\ref{n_far}). To the lowest order in $1 /
\alpha_1$, we get
\begin{equation}
          p(y) = -({\pi} / {16}) \sigma_{x x} a^4(y).
\label{p}
\end{equation}
Next, as in the case of a narrow constriction, we must account for the
renormalization $\sigma \to \sigma + \Delta\sigma$ of the bare
parameters. From Eqs. (\ref{n_far}) and (\ref{p}) we obtain
\begin{equation}
\Delta\sigma(0, y) = \frac{\sigma_{x x}}{32} \int\limits_{-y_0}^{y_0}
\frac{d y^\prime [a^4(y^\prime) - a^4(y)]}
     {|y - y^\prime|^3 + c a^3(y)},
\quad c \sim 1.
\label{Delta_sigma}
\end{equation}
Evaluating this integral and substituting $\Delta\sigma(0, y)$ into
Eq.~(\ref{w_perp}), we find a divergent log-correction to $\alpha = y /
a$. To facilitate the comparison with Eq.~(\ref{b}), the
final result can be presented in the form
\begin{equation}
 b(x) = \alpha_1 |x| \left(
        1 - \frac{12}{\alpha_1^2} \ln \left|\frac{x_0}{x}\right|
\right).
\label{b_wide}
\end{equation}
The most likely behavior consistent with both Eqs.~(\ref{b}) and
(\ref{b_wide}) is as follows. A constriction that appears very wide,
$\alpha = \alpha_1 \gg 1$, at large distances, renormalizes first into
an $\alpha \sim 1$ structure at $x = x_c$,
\begin{equation}
                   x_c \sim x_0 \exp(-\alpha_1^2 / 12),
\label{x_c}
\end{equation}
and then into an adiabatically narrowing small opening-angle
constriction at even smaller $x$. If so, Eq.~(\ref{b}) is a universal
asymptotic law.

Concomitantly, we expect that Eq.~(\ref{Phi_y}) applies only at $|y| \gg x_c$;
otherwise, it is replaced by
\begin{equation}
\Phi(0, y) = \frac{e^2}{\kappa}
\frac{\sigma_{y y}^{3 / 2}}{\sigma_{x x}^{1 / 2}} x_c y^2,
\quad |y| \ll x_c,
\label{Phi_y}
\end{equation}
in a nominal agreement with the popular models~\cite{Fertig_87,Buttiker_90}
for the electrostatic potential near the constriction.

%
% FIG. 2
%
\begin{figure}
\centerline{
\includegraphics[width=2.1in,bb=184 338 400 685]{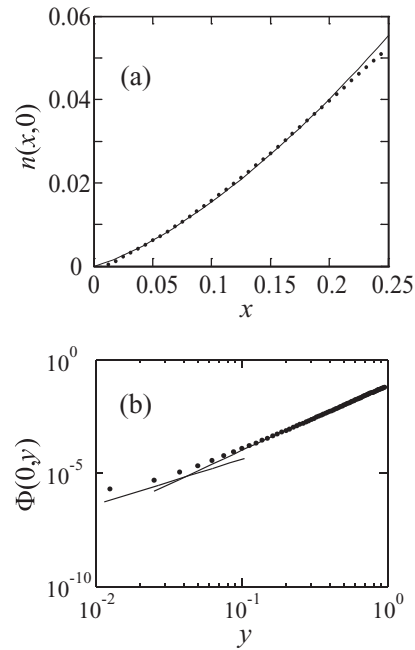}
}
\vspace{0.1in}
%\setlength{\columnwidth}{3.2in}
%\centerline{
\caption{(a) Numerical (dots) and analytical (solid
line) results for $n(x, 0)$ at $U_{x x} = 1$, $U_{y y} = 4$, $U = 2$.
(b) $\Phi(0, y)$ (dots) at $U_{x x} = 1$, $U_{y y} = 0.2$, $U = 3.125$.
The slopes for quadratic and cubic dependences are shown for comparison.
\label{Fig_saddle}
}
%}
\end{figure}

\subsection{Numerical results}

In order to verify the outlined analytical theory we done a series of
numerical simulations. In these simulations the sought $n(x, y)$, the
external potential in the form $\Phi_{ext} = -(1 / 2) U_{x x} x^2 + (1 /
2) U_{y y} y^2 + U x^4$, and the Coulomb interaction kernel were
discretized on a real-space square grid. The units used in the
simulations were $e = \kappa = 1$, and the length unit was the size of
the grid cell. In each run, the energy of the system was minimized
numerically for a trial $\mu_e$ with the help of
\textsc{MATLAB}$^\textsc{TM}$ \textsf{QUADPROG} library function. The
value of $\mu_e$ at which the constriction just opens up was then found
by minimizing the combination $(\Phi - \mu_e)^2 + n^2$ at the point
${\bf r} = (0, 0)$. To avoid a difficult task of reconstructing $b(x)$
from $n$-data defined on the discrete grid, $n(x, 0)$ was studied
instead. According to Eq.~(\ref{n_wire}), $n(x, 0) \propto b(x)$, so
that the scaling form Eq.~(\ref{b}) should equally apply to $n(x, 0)$.
Shown in Fig.~\ref{Fig_saddle}(a) is the numerically calculated $n(x,
0)$ for the $U_{x x} = 1$, $U_{y y} = 4$, the grid size $161 \times 81$,
and the system size $|x| \leq 1 / 2$, $|y| \leq 1 / 4$. On the same
graph we plot the best fit to Eqs.~(\ref{b}) and (\ref{n_wire}), which
is obtained for~\cite{Comment_on_U_yy} $\alpha_0 = 1.15$, $x_0 = 0.82$.
Considering that for the chosen simulation parameters we are not yet
deep inside the asymptotic regime $\alpha \ll 1$ [in the interval of $x$
shown in Fig.~\ref{Fig_saddle}(a), $\alpha$ calculated according to
Eq.~(\ref{n_wire}) ranges from $0.5$ to about $0.2$], the quality of the
fit is quite acceptable. In other words, we think that our computer
simulations do support the validity of Eq.~(\ref{b}).

In Fig.~\ref{Fig_saddle}(b), $\Phi(0, y)$ for the case $U_{x x} / U_{y
y} = 5$ is presented. In agreement with Eq.~(\ref{Phi_y_wide}), it shows
a cubic $y$-dependence at large $y$ and is more consistent with the
quadratic law of Eq.~(\ref{Phi_y}) at small $y$. This again reinforces
our case for universality of Eq.~(\ref{b}).

Concluding this Section, we would like to make a few brief comments on
the relevance of the obtained results for experiments. First, our
predictions for $n(\textbf{r})$ and $\Phi(\textbf{r})$ can be directly
verified by a number of currently available high-resolution imaging
techniques, e.g., near-field optical microscopy,~\cite{Eytan_98}
conductance interferometry,~\cite{LeRoy_02} electrostatic force
microscopy,~\cite{Bachtold_00} or local
potentiometry.~\cite{Ilani_00,Ilani_04} Second, it is feasible that the
differences between the found solution and commonly assumed forms of $n$
and $\Phi$ can also be detected in transport through a quantum point
contact. The signatures of such deviations and the question of how they
could be amplified by a suitable design of the point contact warrant
further study.

%%%%%%%%%%%%%%%%%%%%%%%%%%%%%%%%%%%%%%%%%%%%%%%%%%%%%%%%%%%%%%%%%%%%%%%%%
\section{Checkerboard}
\label{Sec:Chess}

As an application of the obtained results to a more complicated
geometry, in this Section we examine the 2D electron liquid situated
on the periodic charge-density background
\begin{equation}
\sigma = n_e + \frac{n_1}{2} \left(
\cos \frac{2 \pi x}{b_x} + \cos \frac{2 \pi y}{b_y}
\right),\quad b_x \geq b_y.
\label{sigma_cb}
\end{equation}
We refer to this model as the ``checkerboard.'' It is interesting
because it allows one to study the interplay of the three basic building
blocks (a droplet, an isolated DR, and a saddle-point) that exist for a
general $\sigma(\textbf{r})$. The complete analytical solution of the
electrostatic problem for the checkerboard geometry remains however
unknown.~\cite{Comment_on_checkerboard} Instead, we will
present a numerical solution and will discuss how the results of
Secs.~\ref{Sec:Island}--\ref{Sec:Saddle} can be used to understand its
structure. We will also derive some exact asymptotics and finally, at
the end of this Section, we will discuss a semianalytic
\textit{ansatz\/} that reproduces many properties of the numerical
solution with a high accuracy, in particular, its energy as a function
of $n_e$.

\begin{figure}
\centerline{
\includegraphics[width=3.0in,bb=139 256 489 572]{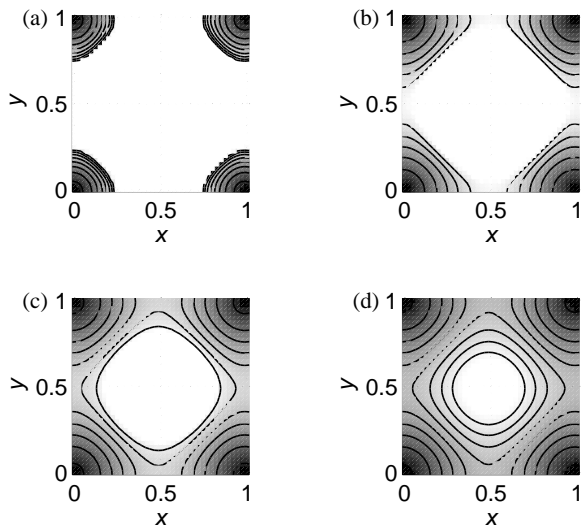}
}
\vspace{0.1in}
%\setlength{\columnwidth}{3.2in}
%\centerline{
\caption{The ground-state density distribution computed numerically on a
$40 \times 40$ square grid for the symmetric checkerboard $b_x = b_y = 1$.
Darker areas correspond to higher $n(\textbf{r})$. The area shown in each
subplot extends beyond the boundary of the unit cell by a half grid cell
in each direction. (a) low density, $n_e = 0.0873 n_1$. (b) $n_e = 0.246
n_1$, which is slightly above $n_p$. (c) $n_e = 0.470 n_1$, about twice
larger than $n_p$ (d) high density, $n_e = 0.836 n_1$. The solid lines
are $n = \text{const}$ contours for a set of linearly spaced densities.
These densities are different in each subplot and are chosen to minimize
uncertainties in the contour positions that arise due to the
discreteness of the grid.
\label{Fig_cb_z1}
}
%}
\end{figure}

We start with numerical results, which are shown in
Fig.~\ref{Fig_cb_z1}. These plots represent the distribution of $n({\bf
r})$ within the unit cell $0 < x \leq b_x$, $0 < y \leq b_y$ that are
computed by a numerical program similar to that described in
Sec.~\ref{Sec:Saddle}. In this particular simulation $b_x = b_y$ so that
the unit cell is a square.

%
% FIG. 3
%

As one can see from Fig.~\ref{Fig_cb_z1}(a), at low density, $n_e \ll
n_1$, the electron liquid is broken into isolated nearly circular
droplets. The droplets surround the maxima of $\sigma(\textbf{r})$ that are
located at the corners of the unit cell. As $n_e$ increases at fixed
$n_1$, the droplets grow. Their boundaries progressively deviate from
the circular form as they become funneled towards the nearest
saddle-points of $\sigma(\textbf{r})$, which are located at the midpoints
of the edges of the unit cell. At some density $n_{p}$ (percolation
point) the droplets merge. In the symmetric checkerboard simulated on
$40 \times 40$ grid, this occurs at the average density of
\begin{equation}
                       n_p = 0.22 n_1,\quad b_x = b_y.
\label{n_p_z1}
\end{equation}
From experiments with different grid sizes, we concluded that the above
value should be close to the percolation threshold in the continuum
limit but no detailed finite-size scaling was attempted.

Figure~\ref{Fig_cb_z1}(b) shows the density profile slightly above $n_p$
where the continuous path through the electron liquid already exists. At
$n_p < n_e < n_1$ the most noticeable change that takes place as $n_e$
continues to increase is the contraction of the depletion hole at the
center of the unit cell, see Figs.~\ref{Fig_cb_z1}(c) and (d). Finally,
at $n_e \geq n_1$ (not shown) the electron liquid becomes free of the
DRs and its profile faithfully repeats the background, $n(\textbf{r}) =
\sigma(\textbf{r})$.

\begin{figure}
\centerline{
\includegraphics[width=2.3in,bb=158 188 416 595]{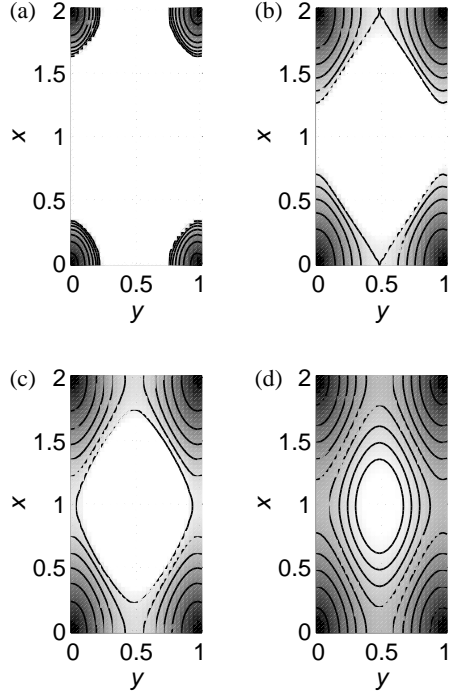}
}
\vspace{0.1in}
%\setlength{\columnwidth}{3.2in}
%\centerline{
\caption{Similar to Fig.~\ref{Fig_cb_z1} but for $b_x = 2$, $b_y = 1$. (a)
low density, $n_e = 0.0549 n_1$. (b) $n_e = 0.269 n_1$, which is between
$n_{p y}$ and $n_p$. (c) $n_e = 0.384 n_1$, which is above $n_p$. (d)
high density, $n_e = 0.897 n_1$.
\label{Fig_cb_z2}
}
%}
\end{figure}

In the asymmetric checkerboard, $b_x > b_y$, the evolution of the ground
state with increasing $n_e$ is similar, except that the transition to
the global percolation takes place in two steps. First, at some density
$n_{p y}$ droplets merge into continuous metallic chains that run
parallel to the $y$-axis. Subsequently, at $n_p > n_{p y}$, the chains
become interconnected. This behavior is illustrated in
Fig.~\ref{Fig_cb_z2} where we display the results of our simulations for
$b_x / b_y = 2$ on the $30 \times 60$ grid. For this grid size, the two
aforementioned thresholds were found to be
\begin{equation}
               n_{p y} = 0.17 n_1,\quad n_p = 0.31 n_1,\quad
               \frac{b_x}{b_y} = 2.
\label{n_p_z2}
\end{equation}
Note that in the asymmetric checkerboard the boundaries of the DRs are
elongated along the $x$-direction. In particular, the small droplets at
low $n_e$ and the small DRs at high $n_e$ are elliptic in shape.

%
% FIG. 4
%

Our goal in the rest of this Section is to develop analytical approaches
that are able to reproduce the above numerical findings.

\subsection{Exact analytical asymptotics}

The structure of the ground state can be determined asymptotically
exactly in the two limits, the low density ($n_e \ll n_1$) and the high
density ($n_1 - n_e \ll n_1$).

We start with the low-density case. Let us split the total charge
background $\sigma(\textbf{r})$ into a part with zero mean, $\sigma({\bf
r}) - n_e$, and a uniform charge density $n_e$. The former produces the
electrostatic potential
\begin{equation}
\Phi_1(\textbf{r}) = -\frac{e n_1 b_x}{2 \kappa}
\left(
\cos \frac{2 \pi x}{b_x} + \frac{b_y}{b_x} \cos \frac{2 \pi y}{b_y}
\right).
\label{Phi_1_cb}
\end{equation}
Function $\Phi_1(\textbf{r})$ has the minima at the corners of the unit
cell and this is the reason why the metallic droplets that form at small
$n_e$ reside there. Each droplet has the electric charge $Q = e n_e b_x
b_y$. Consider the droplet centered at $(0, 0)$ and denote by
$\Phi_{ext}(\textbf{r})$ the total potential felt by the electrons in that
droplet due to all the others and the uniform $n_e$ background. To find
$\Phi_{ext}(\textbf{r})$ we can model the other droplets as point charges
$Q$ arranged in the rectangular lattice. In the leading-order
approximation in the parameter $n_e / n_1$, within the small area
covered by the droplet, $\Phi_{ext}$ is related to $\Phi_1$ as follows:
\begin{equation}
\Phi_{ext}(\textbf{r}) =  \Phi_1(\textbf{r})
+ \frac{Q}{\kappa b_x} M\left(\frac{b_x}{b_y}\right).
\label{Phi_ext_cb}
\end{equation}
Here $M(z)$ is the Madelung constant of the rectangular lattice with
the unit cell of size $1 \times z^{-1}$. $M$ can be easily calculated by
the Ewald's method. For example, one finds that $M(1) =
-3.900264920001955$. To determine the size of the droplet we further
notice that $\Phi_{ext}$ admits the expansion analogous to that in
Eq.~(\ref{Phi_e_elliptic_droplet}),
\begin{eqnarray}
e\Phi_{ext} &=& \mu_0 + \frac12 U_{x x} x^2 + \frac12 U_{y y} y^2
             + O(r^4),
\label{Phi_ext_cb_expansion}\\
\mu_0 &=& \frac{e^2 n_e b_y}{\kappa} M\left(\frac{b_x}{b_y}\right)
 - \frac{e^2 n_1}{2 \kappa} (b_x + b_y),
\label{mu_0}\\
U_{x x} &=& 2 \pi^2 \frac{e^2 n_1}{\kappa b_x},\quad
\frac{U_{y y}}{U_{x x}} = \frac{b_y}{b_x}.
\label{U_xx_cb}
\end{eqnarray}
Substituting these equations into the formulas of Sec.~\ref{Sec:Island},
we find the semiaxes $a$ and $b$ of the droplet to be
\begin{eqnarray}
a &=& b_x \left[\frac{3}{2 \pi^2} \frac{K(k_d) - E(k_d)}{k_d^2}
 \frac{b_y}{b_x} \right]^{1 / 3}
 \left(\frac{n_e}{n_1}\right)^{1 / 3},\quad
\label{a_d_cb}\\
b &=& a \sqrt{1 - k_d^2},
\label{b_d_cb}
\end{eqnarray}
where $k_d$ is the solution of Eq.~(\ref{k_d_equation}) for $U_{x x} /
U_{y y}$ specified by Eq.~(\ref{U_xx_cb}). The depleted area fraction
$f_\text{DR}$ is related to $a$ and $b$ as follows:
\begin{equation}
                  f_\text{DR} = 1 - \frac{\pi a b}{b_x b_y}.
\label{f_DR_d}
\end{equation}
Using the equation of Sec.~\ref{Sec:Island}, we can also calculate the
corrections to the electrochemical potential and the inverse
capacitance, $\delta\mu$ and $\Delta C^{-1}$, respectively, in the
droplet state:
\begin{eqnarray}
\Delta C^{-1} &=& \left[\frac{2 \pi^2}{3}
                        \frac{b_x b_y^2 k_d^2}{K(k_d) - E(k_d)}
                        \right]^{1 / 3}
 \left(\frac{n_1}{n_e}\right)^{1 / 3},\quad
\label{Delta_C_cb}\\
\delta\mu &=& \frac32 \frac{e^2}{\kappa} \Delta C^{-1} n_e
                  + \mu_0.
\label{delta_mu_d_cb}
\end{eqnarray}
Equations~(\ref{a_d_cb})--(\ref{delta_mu_d_cb}) are valid for $(b_x
b_y)^{-1} \ll n_e \ll n_1$. At smaller $n_e$ one expects deviations due
to the discreteness of electrons in each droplet. At larger $n_e$ there
are other kind of deviations, from the nonelliptic shape of the droplets
and their strong mutual interaction.

Let us now switch to the opposite limit of of high density, $\delta n
\equiv n_1 - n_e \ll n_1$. In this case we deal with small depletion
holes that surround the negative minima of $\sigma(\textbf{r})$. Such
minima are located at the centers of the checkerboard cells, e.g., $(b_x
/ 2, b_y / 2)$. Expanding $\sigma(\textbf{r})$ given by
Eq.~(\ref{sigma_cb}) around this point and adhering to the notations of
Eq.~(\ref{sigma_elliptic_hole}), we obtain
\begin{eqnarray}
\sigma_0 &=& n_e - n_1 = -\delta n,
\label{sigma_0_cb}\\
\sigma_{x x} &=& \frac{2 \pi^2 n_1}{b_x^2},\quad
\frac{\sigma_{y y}}{\sigma_{x x}} = \frac{b_x^2}{b_y^2}.
\label{sigma_xx_cb}
\end{eqnarray}
Substituting these expressions into the formulas of Sec.~\ref{Sec:Hole},
we get the semiaxes of the depletion holes $a$ and $b$ to be
\begin{eqnarray}
b &=& \frac{b_y}{\pi} \left[
\frac{(1 + k_h^2) E - (1 - k_h^2) K}{k_h^2 E(k_h)}
\right]^{1 / 2} \left(\frac{\delta n}{n_1}\right)^{1 / 2},\quad
\label{b_h_cb}\\
a &=& \frac{a}{\sqrt{1 - k_h^2}},
\label{a_h_cb}
\end{eqnarray}
where $k_h$ is the solution of Eq.~(\ref{k_h_equation}) for $\sigma_{x x} /
\sigma_{y y}$ specified by Eq.~(\ref{sigma_xx_cb}). For the DR area fraction
we get
\begin{equation}
                  f_\text{DR} = \pi \frac{a b}{b_x b_y},
\label{f_DR_h}
\end{equation}
while for $\delta\mu$ and $\Delta C^{-1}$ we find
\begin{eqnarray}
\Delta C^{-1} &=& \frac{4}{3 \pi} \frac{b_y^2}{b_x^2}
\frac{1 - k_h^2}{E(k_h)}
 \left(\frac{\delta n}{n_1}\right)^{3 / 2},
\label{Delta_C_cb_h}\\
\delta\mu &=& -\frac25 \frac{e^2}{\kappa} \Delta C^{-1} \delta n.
\label{delta_mu_h}
\end{eqnarray}
So far we neglected the interaction among the depletion holes. In
principle, such an interaction, which is a subleading correction of a
dipole-dipole type, can be included perturbatively along the lines of
Sec.~\ref{Sec:Saddle}. However, for all $b_x / b_y$ studied in our
numerical simulations, it was estimated to be a tiny effect at all
densities $n_e$ at which the approximation of DRs by elliptic holes is
still adequate. Therefore, we will not discuss such a refinement.
 
%
% FIG. 5
%
\begin{figure}
\centerline{
\includegraphics[width=2.3in,bb=189 490 418 670]{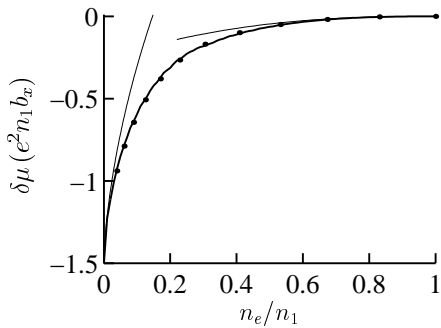}
}
\vspace{0.1in}
%\setlength{\columnwidth}{3.2in}
%\centerline{
\caption{The electrochemical potential correction $\delta\mu$ according
to the analytical asymptotics~[Eqs.~(\protect\ref{delta_mu_d_cb}) and
(\protect\ref{delta_mu_h})] (thin lines), numerical simulations (dots),
and the trial \textit{ansatz\/} method (thick solid line) for the
checkerboard with the unit cell of aspect ratio $b_x / b_y = 2$.
\label{Fig_cb_var_z2}
}
%}
\end{figure}

The comparison between the analytical asymptotics and the numerical data
for $\delta\mu$ is shown in Fig.~\ref{Fig_cb_var_z2} for the case $b_x /
b_y = 2$. As one can see, the droplet picture [Eq.~(\ref{delta_mu_d_cb}),
the left thin line in Fig.~\ref{Fig_cb_var_z2}] remains accurate up to
$n_e \sim 0.07 n_1$. The isolated depletion hole approximation
[Eq.~(\ref{delta_mu_h}), the other thin line in
Fig.~\ref{Fig_cb_var_z2}] is accurate at $n_e \gtrsim 0.4 n_1$. We
conclude that our analytical asymptotics, which are basically the
perturbation theory results, indeed work at low and at high $n_e$, as
expected.

\subsection{Trial ansatz method}

It is also aparent from Fig.~\ref{Fig_cb_var_z2} that the derived
analytical formulas fail at intermediate $n_e$. For example, at the
percolation threshold, $n_p \approx 0.31 n_1$, the actual value of
$\delta\mu$ is about a factor of two off the nearest analytical
asymptote. Going to higher orders in perturbation theory to reduce the
discrepancy appears to be cumbersome and impracticable. It seems that
the quantitatively accurate description of the ground state of the
checkerboard model at intermediate $n_e$ is currently beyond the reach
of controlled analytical methods.

There is however an alternative approach, idea of which was introduced
in Ref.~\onlinecite{Fogler_xxx}. Strictly speaking, this approach is
uncontrolled yet it is semi-analytical and as we will show below, it
reproduces the behavior of $\delta\mu$ at all $n_e$ remarkably well,
both for the symmetric and for the asymmetric checkerboards. In its
simplest implementation, this method amounts to adopting the following
trial \textit{ansatz\/} (TA) for the ground state density distribution:
\begin{equation}
 n_a(\textbf{r}) = \theta(\sigma - \sigma_\text{DR})
                \sqrt{\sigma^2(\textbf{r}) - \sigma_\text{DR}^2}.
\label{n_a}
\end{equation}
Here $\theta(x)$ is the step-function and $\sigma_\text{DR}$ is a constant
that must obey the condition
\begin{equation}
 \int\limits_0^{b_x} {d x}
 \int\limits_0^{b_y} {d y} n_a(x, y) = n_e b_x b_y.
\label{sigma_DR}
\end{equation}
Clearly, $n_a(\textbf{r})$ is entirely fixed by $n_e$ and $\sigma(\textbf{r})$
with no adjustable parameters. Why choosing the trial state in this
form? Several reasons can be given. First, it is consistent with the
notion that $n(\textbf{r})$ is determined primarily by the behavior of
$\sigma$ at points nearby. (After all, the interactions do decay with
distance). Second, unless a function $\sigma(\textbf{r})$ possesses
multiple widely different lengthscales, the behavior of $\sigma$ in the
vicinity of a given point $\textbf{r}$ is dictated predominantly by the
value of $\sigma(\textbf{r})$ at the same point (e.g., small $\sigma$ tend
to be located near minima, large $\sigma$ --- near maxima).
Consequently, the purely local \textit{ansatz\/} $n(\textbf{r}) =
n_a[\sigma(\textbf{r})]$ seems reasonable. Third, Eq.~(\ref{n_a}) preserves
the two asymptotic characteristics of the exact solution: a square-root
singularity at the edges of the metallic regions
(cf.~Sec.~\ref{Sec:Introduction}) and the perfect screening $n \to
\sigma$ at large $n$ (cf.~Sec.~\ref{Sec:Hole}). Fourth, one can verify
that Eq.~(\ref{n_a}) is exact for the DR in the from of an infinite slit
[Eq.~(\ref{n_slit})] and is also rather accurate for the round depletion hole
[Eq.~(\ref{n_round_hole})].

Perhaps, the only serious deficiency of the proposed TA is the omission
of the funneling effect of the saddle points. Indeed, according to
Eq.~(\ref{n_a}) the boundaries of the DRs coincide with the $\sigma({\bf
r}) = \sigma_\text{DR} = \text{const}$ contour, whereas we showed in
Sec.~\ref{Sec:Saddle} that there are logarithmic deviations from such a
behavior, and these are noticeable in Figs.~\ref{Fig_cb_z1}
and~\ref{Fig_cb_z2}. At any rate, the \textit{ansatz\/}~(\ref{n_a}) is
probably the simplest form that one can write down, so it makes sense to
examine how it performs. Having learned its strengths and limitations,
one will be in a better position to apply this kind of methods in
situations where the brute force numerical simulations are difficult,
such as in the models of disordered systems.~\cite{Efros_93}

\begin{figure}
\centerline{
\includegraphics[width=2.4in,bb=190 293 416 660]{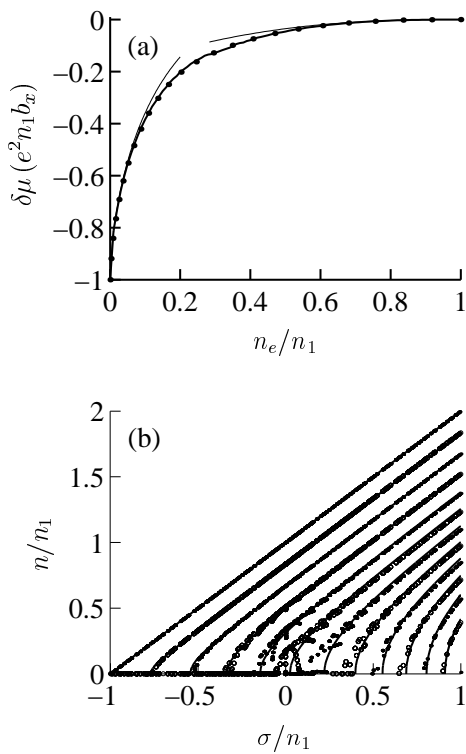}
}
\vspace{0.1in}
%\setlength{\columnwidth}{3.2in}
%\centerline{
\caption{Comparison between the TA method, analytical asymptotics, and
numerical simulations for the symmetric checkerboard, $b_x = b_y$. (a)
The correction $\delta\mu$ [cf.~Eq.~(\protect\ref{delta_mu})] to the
electrochemical potential as a function of $n_e$. The meaning of the
lines and the dots is the same as in Fig.~\ref{Fig_cb_z2}. (b) The
scatter plot of $n$ \textit{vs.} $\sigma$. Solid curves are the
predictions of the TA for same densities $n_e$ as the dots at the top
graph but skipping every other $n_e$ point for clarity. The leftmost curve
is for $n_e = n_1$, the rightmost one (which degenerates into a single
point) --- for $n_e = 0$. The symbols nearby each curve are from
numerical simulations for the corresponding $n_e$.
\label{Fig_cb_var}
}
%}
\end{figure}

The implementation of the TA method goes as follows. First one selects a
reasonably dense set of $n_e$ and determines the corresponding
$\sigma_\text{DR}$ by solving Eq.~(\ref{sigma_DR}) on the computer. In
practice, we did it by approximating the integrals in
Eq.~(\ref{sigma_DR}) by a sum over the grid points. Then, for each
$\sigma_\text{DR}$, one evaluates the total energy energy of the
corresponding trial state~(\ref{n_a}). Finally, the electrochemical
potential $\delta\mu$ is computed by a numerical differentiation of the
total energy with respect to $n_e$. The results of such calculations are
shown by the thick lines in Figs.~\ref{Fig_cb_var_z2} and
\ref{Fig_cb_var}(a). As one can see, the agreement between the TA method
results for $\delta\mu$ and the corresponding numerical datapoints is
very good.

%
% FIG. 6
%

To test the TA method further we can directly compare the density
distribution $n_a(\textbf{r})$ with the numerically determined ground state
$n(\textbf{r})$. Such a comparison is shown in Fig.~\ref{Fig_cb_var}(b)
where we present a scatter plot of $n$ \textit{vs.} $\sigma$, for the
case $b_x = b_y$. The spread of the symbols (numerical data) with
respect to the the solid lines indicates that our TA is certainly not
exact. However, such a spread is not dramatic, and so Eq.~(\ref{n_a}) is
a viable approximation, especially at low and at high $n_e$.

\begin{figure}
\centerline{
\includegraphics[width=2.2in,bb=190 490 410 670]{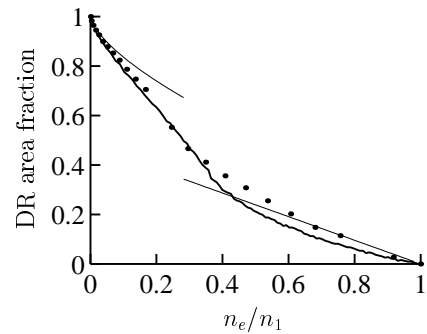}
}
\vspace{0.1in}
%\setlength{\columnwidth}{3.2in}
%\centerline{
\caption{Depleted area fraction according to the analytical
asymptotics of Eqs.~(\ref{f_DR_d}) and~(\ref{f_DR_h})
(thin lines), numerical simulations (dots), and the TA
method (thick line) for the symmetric checkerboard.
\label{Fig_cb_A}
}
%}
\end{figure}

One more quantity we can do the comparison for is the DR area fraction
$f_\text{DR}$. As one can see from Fig.~\ref{Fig_cb_A}, the TA method
performs quite well at all $n_e$, while the analytical asymptotics
[Eqs.~(\ref{f_DR_d}) and~(\ref{f_DR_h})] are obeyed in their respective
validity domains.

%
% FIG. 7
%

Finally, let us discuss the estimate of the percolation threshold
that follows from the TA. According to the TA, the boundaries of the DR
are defined by the equation $\sigma(\textbf{r}) = \sigma_\text{DR}(n_e)$.
Therefore, the percolation occurs at the average density $n_p^\ast$ that
satisfies the relation $\sigma_\text{DR}(n_p^\ast) = n_p^\ast$. Under
this condition, the DR boundary passes simultaneously through all the
saddle-points in the system. For example, within a single rectangular
unit cell, the DR has the shape of a rhombus with vertices at the
midpoints of the cell edges. Solving the above equation numerically, we
found $n_p^\ast \approx 0.31 n_1$. This number is independent of $z_{cb} =
b_y / b_x$ because checkerboards with different $z_{cb}$ can be mapped onto
each other by rescaling the coordinate axes. Within the TA, such a
rescaling does not change the topology of the DRs or the average
electron density.

Clearly, the TA is unable to resolve the existence of two separate
thresholds, $n_{p y}$ and $n_p$ in the asymmetric checherboard. Within
the TA, the percolation occurs simultaneously in the $x$- and
$y$-directions. Still, $n_p^\ast$ is remarkably close to the upper
(global) percolation threshold determined numerically for $z_{cb} = 2$,
see Eq.~(\ref{n_p_z2}). This is also the case at larger $z_{cb}$, e.g.,
at $z_{cb} = 3$ where we found $n_p = 0.32 n_1$. On the other hand, at
lower $z_{cb}$ the discrepancy grows and reaches its largest relative
size of about $30\%$ at $z_{cb} = 1$, see Eq.~(\ref{n_p_z1}). We believe
that these discrepancies (one threshold instead of two and the value of
$n_p$) originate from the two drawbacks of the TA method we mentioned
earlier. One is its unability to handle widely separate lengthscales,
which is the case in checkerboards with large $z_{cb}$. The other is its
weakness in dealing with the saddle-points. The funneling effect of the
saddle-points allows the electron droplets to reach them sooner as $n_e$
increases. Therefore, the continuity of the electron liquid is
established at a lower $n_e$ compared to that predicted by the TA.

One may wonder why the TA method is able to predict $n_p$ with a much
higher accuracy~\cite{Fogler_xxx} ($\sim 10\%$) in the case of a random
$\sigma({\bf r})$. One possible explanation is as follows. The funneling
effect of the saddle-points that is mishandled by the TA method is
especially pronounced in the checkerboard geometry because all the
saddle-points have the same value of $\sigma$, so that the percolation
contour has to pass through all of them simultaneously. Thus, the
inaccuracy of our TA is maximized precisely at $n = n_p$. In contrast,
in the case of a random $\sigma(\textbf{r})$ the percolation contour
passes precisely through the center of a saddle-point very rarely, and
so the TA works very well.

We conclude that the TA method is an excellent and convenient tool for
determining the ground-state energy and depleted area fraction but it
may be less accurate when it comes to more subtle parameters of the
real-space structure, especially if those are heavily dominated by the
saddle-points or a hierarchy of multiple lengthscales.

\acknowledgments

This work is supported by Chris and Warren Hellman Fund and by Alfred P.
Sloan Foundation. I am grateful to Sen Yang for participation in early
stages of this project. I thank I. A. Larkin and B. I. Shklovskii for
valuable comments on the manuscript.

%%%%%%%%%%%%%%%%%%%%%%%%%%%%%%%%%%%%%%%%%%%%%%%%%%%%%%%%%%%%%%%%%%%%%%%%%
\appendix
%%%%%%%%%%%%%%%%%%%%%%%%%%%%%%%%%%%%%%%%%%%%%%%%%%%%%%%%%%%%%%%%%%%%%%%%%

\section{Ellipsoidal harmonics}
\label{Sec:Ellipsoidal}

\begin{widetext}

The ellipsoidal harmonics~\cite{Byerly_book} of the first and the second
kinds, $E_m^p(\xi)$ and $F_m^p(\xi)$, respectively, are defined as the
two linearly independent solutions of the Lam\'e equation (for
$\Lambda$)
\begin{equation}
f(\xi) \frac{d^2 \Lambda}{d \xi^2}
 + \frac12 \frac{d f}{d \xi} \frac{d \Lambda}{d \xi}
 = [m (m + 1) \xi^2 - (1 + k_h^2) p] \Lambda,\quad
f(\xi) \equiv (\xi^2 - 1)(\xi^2 - k_h^2).
\label{Lame_equation}
\end{equation}
For each $m$, which has to be a natural number, $p$ can take any of
$2 m + 1$ different values that depend on $k_h$. Functions
$E_m^p(\lambda)$ and $F_m^p(\lambda)$ at $\lambda^2 \geq 1$ are related
by
\begin{equation}
    F_m^p(\lambda) = (2 m + 1) |E_m^p(\lambda)|
\int\limits_\lambda^\infty
\frac{d l}{\sqrt{f(l)}\, |E_m^p(l)|^2}.
\label{F_from_E}
\end{equation}
At large $\lambda$, $F_m^p(\lambda) \propto 1 / \lambda^{m + 1}$. As a
rule, $F_m^p$'s are not expressed in terms of elementary functions. For
example, for $E_m^p$ given by Eqs.~(\ref{E_1})--(\ref{E_3}),
Eq.~(\ref{F_from_E}) leads to the following $F_m^p$:
\begin{eqnarray}
\frac13 F_1^{p_1}(\lambda) &=& \frac{1}{1 - k_h^2}
\left[\frac{\sqrt{\lambda^2 - k_h^2}}{\lambda} - \sqrt{\lambda^2 - 1}\,
E\Bigl(\arcsin\frac{1}{\lambda}, k_h\Bigr)\right],
\label{F_1}\\
\frac17 F_3^{r_\pm}(\lambda) &=&
\frac{(\lambda^2 - C_\pm) \sqrt{\lambda^2 - k_h^2}}
     {(C_\pm - 1)^2 (1 - k_h^2)}
+ \sqrt{\lambda^2 - 1}\, (\lambda^2 - C_\pm)
\left[A_\pm E\Bigl(\arcsin\frac{1}{\lambda}, k_h\Bigr)
    - B_\pm F\Bigl(\arcsin\frac{1}{\lambda}, k_h\Bigr)\right],
\label{F_3}\\
A_\pm &=& \frac{1 - k_h^2 + 2 C_\pm^2 - 2 C_\pm k_h^2}
               {2 C_\pm (C_\pm - 1)^2 (k_h^2 - C_\pm) (1 - k_h^2)},
\quad
B_\pm = \frac{1}{2 C_\pm (C_\pm - 1) (k_h^2 - C_\pm)},
\label{A_and_B}
\end{eqnarray}
where $E$ and $F$ are the elliptic integrals.~\cite{Gradshteyn_Ryzhik}
However, for $k_h = 0$, these formulas simplify to
\begin{eqnarray}
\frac13 F_1^{p_1}(\lambda) &=& 1 - \arcsin({1}/{\lambda}),
\label{F_1_circle}\\
\frac17 F_1^{r_+}(\lambda) &=& \frac{5}{12} \Bigl[15 \lambda^2 - 11
- 15 \Bigl(\lambda^2 - \frac{2}{5}\Bigr) \sqrt{\lambda^2 - 1}\,
\arcsin({1}/{\lambda}) \Bigr],
\label{F_3_+_circle}\\
\frac17 F_1^{r_-}(\lambda) &=& -\frac{1}{4 \lambda^2}
-\frac{5}{8} + \frac{15}{8} \lambda^2 - \frac{15}{8} \lambda^2
\sqrt{\lambda^2 - 1} \arcsin({1}/{\lambda}).
\end{eqnarray}
\end{widetext}
As an application of these formulas, one can derive the electrostatic
potential $\Phi(\textbf{r}, z)$ around the elliptic DR discussed in
Sec.~\ref{Sec:Hole}. To do so one needs to substitute Eqs.~(\ref{F_1})
and (\ref{F_3}) for $F_m^p$, Eqs.~(\ref{E_1}) and (\ref{E_3}) for
$E_m^p$, and also Eqs.~(\ref{alpha_p_1}) and (\ref{alpha_r_-}) for
$\alpha_m^p$ into the series expansion~(\ref{Phi_hole_expansion}).
Combining such an expression for $\Phi(\textbf{r}, z)$ with
Eq.~(\ref{n_hole_from_Phi_II}) one can then, in principle, deduce the
formula for the density profile $n(\textbf{r})$ of the electron liquid
outside the DR. However, this calculation is not presented here because
for a generic $k_h$ the result is rather unilluminating. The two notable
exceptions are $k_h = 0$, where one obtains a circular DR with $n(\textbf{r})$
given by Eq.~(\ref{n_round_hole}), and $k_h = 1$ where the DR is an
infinite depletion strip and Eq.~(\ref{n_slit}) holds.

%%%%%%%%%%%%%%%%%%%%%%%%%%%%%%%%%%%%%%%%%%%%%%%%%%%%%%%%%%%%%%%%%%%%%%%%%


\begin{thebibliography}{99}
\bibitem{Comment_on_Dyson} This proof is sketched in
Ref.~\onlinecite{Sneddon_book} without reference to Dyson.
\bibitem{Comment_on_cutoff}As already explained above, in a
physical system there is an additional small-distance cutoff equal
to $a_\text{e-e}(x)$.
\bibitem{Comment_on_U_yy} The renormalized value $U_{y y}^\text{ren}
\approx 3.60$ also needed here was found by fitting numerical $\Phi(0,
y)$ to a parabola. As expected, $U_{y y}^\text{ren}$ is a bit larger
than its perturbative estimate $U_{y y}^\text{ren} = U_{y y} - (U_{x x}
/ 2) = 3.5$.
\end{thebibliography}
\end{document}